\documentclass[preprint,aps,tightenlines,showpacs,showkeys,amsmath,amssymb]{revtex4}
\usepackage{graphicx}
\usepackage{dcolumn}
\usepackage{bm}
\usepackage{amsmath} 

\begin{document}

\title{Thomas-Fermi theory for atomic nuclei revisited}

\author{M. Centelles$^1$\footnote{{\em E-mail address:} mario@ecm.ub.es},
P. Schuck$^2$, and X. Vi\~nas$^1$}

\affiliation{
$^1${\it Departament d'Estructura i Constituents de la Mat\`eria,
Facultat de F\'{\i}sica,
\\
Universitat de Barcelona,
Diagonal {\sl 647}, {\sl 08028} Barcelona, Spain}
\\
$^2${\it Institut de Physique Nucl\'eaire, IN{\sl 2}P{\sl 3}--CNRS,
\\
Universit\'e Paris--Sud, {\sl 91406} Orsay-C\'edex, France}
}


\begin{abstract}
The recently developed semiclassical variational Wigner-Kirkwood (VWK)
approach is applied to finite nuclei using external potentials and
self-consistent mean fields derived from Skyrme interactions and from
relativistic mean field theory. VWK consists of the Thomas-Fermi part
plus a pure, perturbative $\hbar^2$ correction. In external
potentials, VWK passes through the average of the quantal values of
the accumulated level density and total energy as a function of the
Fermi energy. However, there is a problem of overbinding when the
energy per particle is displayed as a function of the particle number.
The situation is analyzed comparing spherical and deformed harmonic
oscillator potentials. In the self-consistent case, we show for Skyrme
forces that VWK binding energies are very close to those obtained from
extended Thomas-Fermi functionals of $\hbar^4$ order, pointing to the
rapid convergence of the VWK theory. This satisfying result, however,
does not cure the overbinding problem, i.e., the semiclassical energies
show more binding than they should. This feature is more pronounced in
the case of Skyrme forces than with the relativistic mean field
approach. However, even in the latter case the shell correction energy
for e.g. $^{208}$Pb turns out to be only $\sim -6$ MeV what is about a
factor two or three off the generally accepted value. As an ad hoc
remedy, increasing the kinetic energy by 2.5\%, leads to shell
correction energies well acceptable throughout the periodic table. The
general importance of the present studies for other finite Fermi
systems, self-bound or in external potentials, is pointed out.
\end{abstract}



\maketitle

\section{Introduction}

One of the most important problems of finite fermion systems such as
nuclei, atoms, helium- and metallic-clusters, quantum dots, etc., is
the determination of the ground-state binding energy and the
corresponding particle density distributions. In the nuclear case, to
overcome the problems encountered when starting from realistic bare
nucleon-nucleon forces, approximate and phenomenological schemes have widely 
been employed. This is the case of the very successful density dependent
Hartree-Fock method with Skyrme \cite{Sky} or Gogny \cite{Gog} forces
in the non-relativistic framework and of the relativistic mean field
theory (non-linear $\sigma-\omega$ model) \cite{SW} in the
relativistic formulation.

To investigate how properties of global character vary with the number
of nucleons $A$, which is the subject of the present work,
semiclassical or statistical techniques are very useful. The best
known example is the nuclear mass formula, based on the liquid drop or
droplet model \cite{Myers}. The success of the mass formula in
describing binding energies lies in the fact that the quantal effects,
i.e. shell corrections, are small as compared with the part of the
energy which smoothly varies with $A$. The perturbative treatment of
the shell correction energy in finite Fermi systems was elaborated by
Strutinsky in the case of nuclei \cite{Strut}. It was proposed to
divide the total quantal ground state energy in two parts:
\begin{equation}
E = \tilde{E} + \delta E.
\label{eq1} \end{equation}
The by far largest part, $\tilde{E}$, varies smoothly with the number
of fermions and is to be associated with the liquid drop energy. It
can be calculated from e.g. the Hartree-Fock (HF) approach using the
Strutinsky smoothing method \cite{Strut}, which is a well defined
mathematical procedure to erase the quantal oscillations in a finite
Fermi system. However, this method may in general be more difficult to
handle than the solution of the full quantal problem if realistic
potentials are used. Thus, the search of alternative methods is an
interesting and still partly open problem, as we will see.
Semiclassical methods of the Thomas-Fermi (TF) type, which evaluate
the smooth part of the energy, have widely been used in atomic,
nuclear and metallic clusters physics. These TF methods, like the
liquid droplet or Strutinsky calculations, smooth the quantal shell
effects and estimate the average part of the HF energy~\cite{BP,Swia}.

The semiclassical methods of the TF type are usually based on the
Wigner-Kirkwood (WK) expansion of the density matrix \cite{WK}. In
this approach, the single-particle density $\rho$ and the kinetic
energy density $\tau$ are expressed by means of functionals of the
one-body single-particle mean field potential $V$. The $\hbar^2$ or
$\hbar^4$  corrections to the lowest-order TF term contain gradients
of $V$ of second or fourth order that arise from the non-conmutativity
between  the momentum $\hat{{\bf p}}$ and position $\hat{{\bf R}}$
operators.  The $\hbar$ corrections to the pure TF particle or kinetic
energy  densities are known to diverge at the classical turning point.
They  are to be considered rather as distributions than as functions
\cite{RS,KCM}, in the sense that only integrated quantities have a real
physical meaning. It has been shown that, in the case of a harmonic
oscillator potential well, the WK theory including up to $\hbar^4$
corrections is equivalent to the Strutinsky average \cite{BP}.

An important property of the WK expansion of the energy in powers of
$\hbar$ concerns its variational content. For a set of non-interacting 
fermions submitted to an external potential, as for instance harmonic
oscillator or Woods-Saxon wells, the variational solution for the
particle density which minimizes the semiclassical WK energy at each
order of the $\hbar$ expansion, is just the WK expansion of the
particle density $\rho$ at the same order in $\hbar$ \cite{SV,CVDSE}.
The method for solving this variational problem \cite{SV,CVDSE}, which
sorts out properly the different powers of $\hbar$ at each step of the
minimization, was called variational Wigner-Kirkwood (VWK) theory.
This VWK method has been applied to describe half infinite nuclear
matter in the self consistent case using Skyrme forces \cite{CVDSE}
and the relativistic mean field approximation \cite{ECV}. The formal
VWK approach, up to $\hbar^4$ order, was developed in \cite{CVDSE}.

Another related approach widely used for
dealing with the semiclassical approximation of the self consistent HF
problem, is based on the so-called density functional theory (DFT).
 The theoretical justification of DFT is
formally provided by the Hohenberg and Kohn theorem \cite{HK}. In the
nuclear context it states that the energy of a set of interacting
nucleons is a unique functional of the local density alone, that is
\begin{equation}
E = \int d \vec{r} \varepsilon \big[\rho (\vec{r})\big].
\label{eq2} \end{equation}
 which reaches its minimal value when calculated with the exact
ground-state  density.
The ground-state density $\rho(\vec{r})$ is determined by a variational 
approach of Eq.(\ref{eq2}) with the constraint of a fixed number of 
particles:
\begin{equation}
\frac{\delta}{\delta \rho} \int d \vec{r} 
\{ \varepsilon \big[ \rho(\vec{r}) 
\big]  - \mu \rho(\vec{r}) \} = 0.
\label{eq3} \end{equation}

In spite of the appeal of Eqs. (\ref{eq2}) and (\ref{eq3}), in general
the exact energy functional is unknown and approximate techniques have
to be worked out. The most popular and successful semiclassical
approach based on DFT and developed together with the use of the
Skyrme forces is the extended Thomas-Fermi (ETF) method. There the WK
$\hbar$ expansion of the density is inverted to recast the kinetic
energy density as a functional $\tau[\rho]$ of the local density and
its derivatives \cite{GV,BGH}. If the potential part of the energy
density is also a known functional of $\rho(\vec{r})$, as it happens
for the Skyrme forces, the approximate energy density functional can
be minimized to obtain an Euler-Lagrange equation like in Eq.(\ref{eq3}).
The solution of this equation will provide the ground-state particle
density and energy. The quantum shell oscillations are absent in the
ETF model, which yields average densities and energies 
with good accuracy \cite{BGH,CPVGB,PS,Li,BB,CVBS,CV}.

Using the VWK method, we have studied in Refs.\cite{SV,CVDSE} the
surface energy of a half infinite Fermi gas embedded in an external
Woods-Saxon potential well. When $\hbar^2$ corrections are taken into
account, the VWK surface energy reproduces the quantal values within
1\% and the agreement is almost perfect when $\hbar^4$ corrections are
considered. This result indicates that quantal Friedel oscillations
have a negligible influence on the nuclear surface energy. We also
have  solved this problem using ETF (that in the case of the external
potential reduces to the use of the ETF kinetic energy density).
However, in this case we find discrepancies between ETF and quantal
surface energies of 10\% and 7\% considering $\hbar^2$ and $\hbar^4$
corrections, respectively. This is an indication that ETF results are
less well converged than the VWK ones. As it also will be discussed
later on, this is mainly due to the fact that VWK properly sorts out
the different orders in $\hbar$ which is not the case in ETF.

In Ref.\cite{CVDSE} we have analyzed the surface energy in
self consistent problems using the TF, VWK, and ETF (up to $\hbar^2$
order) semiclassical approaches in comparison with the quantal (HF)
results. In this study we considered several Skyrme forces that cover
a wide range of effective mass values and incompressibility moduli in
bulk matter at saturation. In general the VWK2 surface energies are
closer to the quantal values than the TF ones. (We call VWK2 the VWK
theory developed up to order $\hbar^2$, similarly VWK4 for the theory
up to order $\hbar^4$, and analogously for the ETF theory.) The ETF2
surface energies are systematically smaller than the corresponding
quantal values, and their absolute error is always larger than in the
VWK2 case. An analysis of VWK and ETF calculations of half infinite
nuclear matter in relativistic mean field theory yields the same kind
of conclusions and the quality of the VWK2 results is seen to be
systematically better than in ETF2 \cite{ECV}.
However, if one compares the situation in the case of self consistent
mean fields with respect to the case of the external Woods-Saxon
potential well, the agreement between the semiclassical and quantal
calculations of the surface energy worsens in the self consistent
case,  pointing to the fact that in the self consistent problems the
semiclassical  $\hbar$-expansions are more involved than in the case
of an external potential. This aspect will be treated with more detail
along this paper.

Summarizing, from the previous discussion it is clear that the VWK and
ETF methods exhibit small but significant differences
\cite{SV,CVDSE,ECV}. The reason lies in the fact, as already
mentioned, that ETF does not properly sort out the different powers in
$\hbar$ and that it partially sums terms to all orders in $\hbar$. On
the other hand, our previous findings in the study of the surface
energy in the self consistent case \cite{SV,CVDSE,ECV} indicate that
the splitting of the quantal binding energies into their smooth and
fluctuating parts is not so well under control (both in ETF and in
VWK) as in the case of an external potential.

The main purpose of this paper is to develop and apply the VWK theory
to {\it finite} nuclei in the self consistent case using both
non-relativistic Skyrme forces and relativistic mean field (RMF)
interactions \cite{SW}. In the next section we present the basics of
the VWK method in the non-relativistic case with an alternative
derivation to the one used in Refs. \cite{SV,CVDSE} to explicitly show
the differences between the VWK and ETF methods. In section III we
first  discuss within WK the external potential case to set the stage
for the study of finite nuclei with self consistent mean fields later.
We show that for strongly triaxially deformed mean field potentials
with absence of any degeneracies, the semiclassical energies are
extremely close to the quantal ones. Approaching sphericity in a
homothetic way the shell structure becomes more and more apparent.
Spherical open shell nuclei are slightly less bound quantally than
semiclassically. This gives rise for the binding energy per particle
as a function of particle number to the typical quantal arch structure
between magic numbers whereas the semiclassical curve is, of course,
monotonous. Section IV is devoted to the self consistent problem in
finite nuclei. First, in section IV.A, we will show that using Skyrme
forces, the VWK2 and ETF4 approaches practically give the same energy
along the periodic table and that this fact is independent of the
Skyrme interaction chosen. This a priori satisfying feature reveals,
however, that the semiclassical approaches VWK2 and ETF4 overbind in
excess, since even for the doubly magic nucleus $^{40}$Ca the
Hartree-Fock results gives less binding than the semiclassical ones.
This is shown and discussed in section IV.B, where also an ad hoc
remedy to this problem is proposed. Several technical aspects are
discussed in Appendices 1 and 2.

\section{The variational Wigner-Kirkwood theory}

The VWK theory has formally been introduced in Refs.
\cite{SV,CVDSE,ECV}. Here we present, for a non-relativistic Skyrme
force, a shortcut derivation in order to show explicitly the
similarities and differences with the ETF method. For the sake of
simplicity we consider symmetric and uncharged nuclei for the moment
and, in this section, a Skyrme force with an effective mass equal the
physical one. In this case the total energy of a nucleus in the ETF
approach up to order $\hbar^2$ (ETF2) is written as \cite{BGH,CPVGB}:
\begin{eqnarray}
E = \int d \vec{r} \{ \frac{\hbar^2}{2 m} \big[ \frac{3}{5} \big(\frac{3 
\pi^2}{2}\big)^{2/3} \rho^{5/3} 
+ \frac{1}{36} \frac{(\nabla \rho)^2}{\rho}
+\frac{1}{3} \Delta \rho \big] 
+ a_0 \rho^2 + a_3 \rho^{2 + \alpha} + a_{12} \, (\nabla \rho)^2 \},
\label{eq4}  \end{eqnarray}
where $a_0= 3 t_0 / 8$, $a_{12}= (9t_1 - 5t_2 - 4t_2 x_2)/64$, $a_3=
t_3 / 16$, and $\alpha$ are the parameters that characterize the
Skyrme interaction.
The terms inside square brackets correspond to the ETF kinetic energy
density $\tau_{ETF}$ up to $\hbar^2$ order. The $\rho^{5/3}$ term is
the well-known pure TF contribution, which is of order $\hbar^0$ in
the  expansion of the kinetic energy density in $\hbar$ powers. The
two remaining terms are of $\hbar^2$ order, and the first one is the
so-called Weizs\"acker term. Clearly, $\tau_{ETF}$ is a functional of
the local density where the gradient terms are of second order in $\hbar$. 

Starting from Eq.(\ref{eq4}) the Euler-Lagrange equation for the local 
density constrained to give $A$ nucleons reads: 
\begin{eqnarray}
\frac{\hbar^2}{2 m} \bigg[ \bigg(\frac{3
\pi^2}{2}\bigg)^{2/3} \rho^{2/3} 
+ \frac{1}{36} \frac{(\nabla \rho)^2}{\rho^2}
- \frac{1}{18} \frac{\Delta \rho}{\rho} \bigg] 
+ 2 a_0 \rho +(2 + \alpha) a_3
\rho^{1 + \alpha} - 2 a_{12} \, \Delta \rho = \mu,
\label{eq5}  \end{eqnarray}
where the chemical potential $\mu$ is the Lagrange multiplier that
ensures the right normalization of the local density $\rho$. In the
ETF method the variational equation (\ref{eq5}) is solved numerically,
for instance using the imaginary-time step method \cite{CPVGB}.
However, the ETF approximation has some consistency problems with
respect to the correct sorting out of powers in $\hbar$ \cite{CVDSE}.
The reason is that the solution of Eq. (\ref{eq5}) contains $\hbar$ at
all orders due to the fact that the Weizs\"acker term in
Eq.(\ref{eq4}) is of order $\hbar^2$. Actually Eq.(\ref{eq5}) has a
similar structure as a Schr\"odinger equation for $\rho$ and thus the
density contains $\hbar$ as an essential singularity \cite{SV}.

In order to properly sort out the different powers in $\hbar$ (to
second order in the present example) one should split the local
density and chemical potential entering in Eq.(\ref{eq5}) into their
$\hbar^0$ and $\hbar^2$ parts:
\begin{equation}
 \rho= \rho_0 + \hbar^2 {\rho_2}
\label{eq6} \end{equation}
and 
\begin{equation}
\mu = \mu_0 + \hbar^2 \mu_2.
\label{eq7} \end{equation}
Using (\ref{eq6}) and (\ref{eq7}), the Euler-Lagrange equation
(\ref{eq5})  can be sorted into $\hbar^0$ and $\hbar^2$ terms. One key
point in the VWK theory is that the minimization is performed for each
order in the expansion parameter $\hbar^2$ separately since, in principle, 
$\hbar$ can be considered as an arbitrary parameter (see Refs.
\cite{SV,CVDSE,ECV} for a more detailed discussion of this point).
Thus from the Euler-Lagrange equation (\ref{eq5}) one obtains
\begin{equation}
 \frac{\hbar^2}{2 m} \big(\frac{3
\pi^2}{2}\big)^{2/3} \rho_0^{2/3} 
+ 2 a_0 \rho_0 +
(2 + \alpha) a_3
\rho_0^{1 + \alpha} - 2 a_{12} \, \Delta \rho_0 - \mu_0 =  0
\label{eq8} \end{equation}
at TF ($\hbar^0$) order, and
\begin{eqnarray}
&& \frac{\hbar^2}{2 m} \big[ \frac{2}{3} \big(\frac{3
\pi^2}{2}\big)^{2/3} \rho_0^{-1/3} {\rho_2} 
 + \frac{1}{36} \frac{(\nabla \rho_0)^2}{\rho_0^2} 
 -\frac{1}{18} \frac{\Delta 
\rho_0}{\rho_0}
 \big] \nonumber \\
&+& 2 a_0 {\rho_2} +
(2 + \alpha)(1 + \alpha) a_3
\rho_0^{\alpha} {\rho_2} - 2 a_{12} \, \Delta {\rho_2} - \mu_2  = 0
\label{eq9}  \end{eqnarray}
for the linearized second order correction.
Another important point in the VWK theory is that the TF local 
density $\rho_0$, i.e. the variational solution of Eq.(\ref{eq8}), 
fulfills the normalization condition:
\begin{equation}
\int d\vec{r} \rho_0 = A
\label{eq10} \end{equation}
and due to the fact that
\begin{equation}
\int d \vec{r} \big(\rho_0 + \hbar^2 {\rho_2} \big) = A,
\label{eq11} \end{equation}
we immediately see that the integral over the second-order density
vanishes. This condition can be assured in adjusting $\mu_2$.

Now, splitting the total energy $E$ into its $\hbar^0$ and $\hbar^2$
contributions and using Eqs. (\ref{eq8}), (\ref{eq10}) and
(\ref{eq11}), one finds that in the VWK approach the energy of
a finite nucleus including corrections of order $\hbar^2$ can be
written as:
\begin{eqnarray}
E = \int d \vec{r} \{ \frac{\hbar^2}{2 m} \big[ \frac{3}{5} \big(\frac{3
\pi^2}{2}\big)^{2/3} \rho_0^{5/3} + \frac{1}{36} \frac{(\nabla 
\rho_0)^2}{\rho_0}
+\frac{1}{3} \Delta \rho_0 \big]
+ a_0 \rho_0^2 +  a_3
\rho_0^{2 + \alpha} + a_{12} \, (\nabla \rho_0)^2 \}.
\label{eq12}  \end{eqnarray}
Thus we arrive at the important result that the total energy up to
order $\hbar^2$ is computed using only the lowest-order solution (TF)
of the Euler-Lagrange equation. In practice this amounts to take the
expression of the total energy as formally given by ETF2 but to
compute it with the TF density solution. Consequently, the VWK
procedure is consistent with the spirit of perturbation theory, since
to calculate the energy at order $\hbar^2$ only requires knowledge of
the solution of $\rho$ to the previous ($\hbar^0$) order
\cite{SV,CVDSE,ECV}. The integral in Eq.\ (\ref{eq12}) is defined
between $r=0$ and the classical turning point $r_t$ where the TF
density $\rho_0$ vanishes. The analysis of Eq.\ (\ref{eq12}) near
$r_t$ shows that $\rho_0$ behaves as $(r_t-r)^2$ and, as a
consequence, the integrand of (\ref{eq12}) is always finite in the
whole domain of definition.

Of course the procedure can be continued to obtain the fourth order
correction, see Ref.\cite{CVDSE} where this has been worked out in a
slightly different way. The fourth order is, however, much more
complicated, and necessitates for instance the knowledge of
${\rho_2}$ which may not easily be accessible \cite{CVDSE}. We
remark that ${\rho_2}$ is not needed in VWK2 for the calculation
of the energy. This is a consequence of the fact that the {\it total}
energy is just the quantity that is minimized and then the use of the
Euler-Lagrange equations allows to eliminate ${\rho_2}$ in
the expansion of the energy. However, the evaluation of other
quantities that are not minimized, e.g. kinetic energies or root mean
square radii, etc., needs the explicit knowledge of ${\rho_2}$
when computed to $\hbar^2$ order.

It should be pointed out that in the general realistic case with
effective mass different from the bare nucleon mass, inclusion of the
spin-orbit potential, etc., the VWK2 method follows the same principle
as in our schematic example. In practice one can take the
corresponding ETF2 expression for the ground-state energy and replace
$\rho(\vec{r})$ by its TF solution $\rho_0(\vec{r})$ which is the
self consistent solution of the lowest-order variational TF equation.
We refer the reader to Eqs. (A1)-(A5) of Ref. \cite{CEV} for the
detailed ETF2 expression of the energy in the case of realistic Skyrme
forces. 

The same VWK theory can be applied to finite range effective forces
such as the Gogny interaction \cite{Gog} although this case will not be
treated explicitly in this paper. For this type of forces the
semiclassical single-particle potential is not only position but also
momentum dependent because of the finite range \cite{CVDSE}. Then, in
addition to the kinetic and spin-orbit $\hbar^2$ corrections to the
energy, there is another $\hbar^2$ contribution coming from the
exchange term. Due to the $k$-dependence of the single-particle
potential, the effective mass also becomes momentum dependent, which
introduces extra terms in the $\hbar^2$ energy not present in the case
of local forces as the Skyrme ones. The reader can find in Ref.
\cite{SV2000} a detailed discussion of the ETF approach in the case of
a general finite-range effective force. In particular, the kinetic and
exchange energy densities in this case are given by Eqs.(39) and (40)
of that reference. On the other hand, we also will consider in this paper
the VWK approach applied to the relativistic mean field theory for the
description of nuclei. The relativistic model automatically contains
the finite range, spin-orbit and density dependence of the
nucleon-nucleon  interaction. The basic relativistic VWK theory up to
$\hbar^2$ order has been worked out in Ref. \cite{ECV} and
applied to the analysis of half infinite nuclear matter. In the case
of finite nuclei the basic equations to be used for the VWK2
calculations are Eqs.(A7)-(A11) of Ref. \cite{CEV} together with
Eqs.(5.8)-(5.12) of Ref. \cite{CVBS} computed with the solution of the
relativistic TF equations.

\section{The external potential case}

In order to get a deeper insight into the behavior of the
semiclassical energies as compared with the quantal ones before the
study of finite nuclei with the use of self consistent mean field
potentials, we first analyze the simpler problem of a set of
non-interacting fermions submitted to an external potential well. In
this case the VWK solution up to $\hbar^2$ order is just the WK expansion
of the local density \cite{SV,CVDSE} as pointed out in the Introduction.
We will consider the model problems of harmonic oscillator and
Woods-Saxon potentials. The discussion of the harmonic oscillator,
apart of being of interest by itself as it is one of the most
important model potentials in quantum mechanics, is relevant in
different areas of physics beyond the context of atomic nuclei, such
as confined electron systems or trapped ultracold fermion gases. A
separate study for the harmonic oscillator potential including
deformation degrees of freedom and the problem of a cavity with sharp
boundaries, where the WK expansion cannot be applied, will be
presented in a forthcoming publication \cite{Leboeuf06}.

One important quantity is the number of states (accumulated level
density) up to an energy $\varepsilon$, which is defined as \cite{RS}
\begin{equation}
N(\varepsilon) = \int^{\varepsilon}_0 g(\varepsilon') d\varepsilon'.
\label{eq13} \end{equation}
The level density $g(\varepsilon)$ is given by
\begin{eqnarray}
g(\varepsilon) &=& {\rm Tr}\/ [\delta (\varepsilon - {\hat H})] =
\frac{\partial}{\partial \varepsilon}
{\cal{L}}^{-1}_{\beta \to \varepsilon} \bigg[ \frac{2}{(2 \pi \hbar)^3} \int 
\int
\frac{C^{\beta}(\vec{r},\vec{p})}{\beta} d\vec{r} d\vec{p} \bigg] \nonumber \\
&=& \frac{2}{(2 \pi \hbar)^3} \int \int
\frac{\partial f_{\varepsilon}(\vec{r},\vec{p})}{\partial \varepsilon} 
d\vec{r} d\vec{p}
\label{eq14} \end{eqnarray}
where ${\cal{L}}^{-1}_{\beta \to \varepsilon}$ is the inverse Laplace
transform  and the factor 2 takes into account spin degeneracy. We
use the notation $C^{\beta}(\vec{r},\vec{p})$ for the Wigner transform
of the single-particle propagator $\hat{C}^{\beta}= \exp{(- \beta
\hat{H})}$,  and $f_{\varepsilon} (\vec{r},\vec{p})$ is the
corresponding Wigner function whose semiclassical expansion up to
order $\hbar^2$ reads \cite{RS}
\begin{eqnarray}
f_{\varepsilon}(\vec{r},\vec{p}) &=& \Theta(\varepsilon - H_{\rm w}) - 
\frac{\hbar^2}{8m} \Delta V \delta' (\varepsilon - H_{\rm w}) \nonumber \\
&+& \frac{\hbar^2}{24m} \big[ (\nabla V)^2 + \frac{1}{m} (\vec{p}
\cdot \nabla)^2 
V \big] \delta'' (\varepsilon - H_{\rm w}), 
\label{eq15} \end{eqnarray}
where $H_{\rm w}$ is the classical mean field Hamiltonian (Wigner
transform of ${\hat H}$).

Inserting Eq.(\ref{eq14}) into (\ref{eq13}) one obtains the
accumulated level density from the Wigner function as
\begin{equation}
N(\varepsilon) = 
\frac{2}{(2 \pi \hbar)^3} \int \int
 f_{\varepsilon}(\vec{r},\vec{p})d\vec{r} d\vec{p}.
\label{eq16} \end{equation}
In the same way the energy of a set of fermions in a potential well
filled up to the Fermi energy $\varepsilon$ can be expressed as
\begin{equation}
E(\varepsilon) = \int^{\varepsilon}_0 \varepsilon' 
g(\varepsilon') d\varepsilon' =
  \frac{2}{(2 \pi \hbar)^3} \int \int
f_{\varepsilon}(\vec{r},\vec{p})H_{\rm w} d\vec{r} d\vec{p} .
\label{eq17} \end{equation}

To simplify the calculation of $N(\varepsilon)$ and $E(\varepsilon)$,
it is helpful to realize that $H_{\rm w}$ is the natural variable for
$f_{\varepsilon}(\vec{r},\vec{p})$. In particular, the classical
spherical  harmonic oscillator (HO) Hamiltonian $H_{\rm w} = p^2/2m + m
\omega^2 r^2/2 = P^2 + Q^2$ can be seen as the square of a radial
component $\sqrt{P^2 + Q^2}$ in polar coordinates, with a polar angle
$\theta = \arctan (P/Q)$. In a similar way, for a more general
potential with spherical symmetry, radial and polar angle coordinates
can be defined in phase space by
\begin{equation}
\sqrt{\tilde{H}_{\rm w}} = \sqrt{H_{\rm w} - V(0)},
\label{eq18} \end{equation}
where $V(0)$ is the bottom of the potential, and
\begin{equation}
\frac{p^2}{2m} = \tilde{H}_{\rm w} \sin^2 \theta, \qquad 
V(\vec{r}) - V(0) = \tilde{H}_{\rm w} \cos^2 \theta .
\label{eq19} \end{equation}
This allows switching from the variables $(r,p)$ to the new ones
$(H_{\rm w},\theta)$ in the integrals over phase space. An advantage of this
procedure is that one automatically circumvents the divergency problems
usually encountered at the classical turning point when the $\hbar^2$
corrections are taken into account. We will use this method to obtain
the results for the accumulated level density and energy as a
function of $\varepsilon$ for an external Woods-Saxon potential that
we will discuss later in this Section.

In the case of an external potential of HO type the integration of
Eqs.(\ref{eq16}) and (\ref{eq17}) can be done analytically. The
semiclassical expressions of the accumulated level density and energy
read 
\begin{equation}
N_{\rm WK}(\varepsilon) = \frac{1}{3} \bigg(\frac{\varepsilon}{\hbar 
\omega}\bigg)^3
- \frac{1}{4} \frac{\varepsilon}{\hbar \omega}
\label{eq20} \end{equation}
and
\begin{equation}
E_{\rm WK}(\varepsilon) = \bigg[ \frac{1}{4} 
\bigg(\frac{\varepsilon}{\hbar \omega}\bigg)^4
- \frac{1}{8} \bigg(\frac{\varepsilon}{\hbar \omega}\bigg)^2 
- \frac{17}{960} \bigg] \hbar \omega,
\label{eq21} \end{equation}
respectively, where the contribution $-17\hbar \omega/960$ in the last
equation comes from the $\hbar^4$ correction. Notice that in a HO
potential there is no $\hbar^4$ correction in $N_{\rm WK}$ \cite{BP}.

For the HO potential the quantal level density can also be
obtained analytically \cite{BB,BJ} and reads:
\begin{equation}
g(\varepsilon) = \frac{1}{\hbar \omega} 
\bigg[ \bigg(\frac{\varepsilon}{\hbar 
\omega}\bigg)^2  - \frac{1}{4} \bigg]
\bigg( 1 + 2 \sum_{M=1}^{\infty} (-1)^M \cos \bigg( 2 \pi M
\frac{\varepsilon}{\hbar \omega} \bigg) \bigg),
\label{eq22} \end{equation} 
which is seen to split into a part that smoothly varies with
$\varepsilon$  and a fluctuating part.
The smooth part is equal to the semiclassical WK expansion of the
level density up to $\hbar^2$ (as already mentioned, the contributions
of higher order in $\hbar$ vanish for the HO potential). The
fluctuating part corresponds to the shell correction and contains all
the quantal effects not included in the WK expansion. The quantal
expressions for the accumulated level density and energy can easily be 
calculated starting from Eq.(\ref{eq22}):
\begin{eqnarray}
N(\varepsilon) &=& N_{\rm WK}(\varepsilon) +
 2 \sum_{M=1}^{\infty} (-1)^M \bigg[\frac{1}{4 \pi^2 
M^2}\frac{\varepsilon}{\hbar \omega}
\cos \bigg(2 \pi M \frac{\varepsilon}{\hbar \omega} \bigg) \nonumber \\
&+& \bigg( \frac{1}{2 \pi M} \bigg(\frac{\varepsilon}{\hbar \omega}\bigg)^2 - 
\frac{1}{4 \pi^3 M^3} -
\frac{1}{8 \pi M}\bigg) \sin \bigg(2 \pi M \frac{\varepsilon}{\hbar \omega}
\bigg) \bigg] \label{eq23} \end{eqnarray}
and
\begin{eqnarray}
E(\varepsilon)
&=& E_{\rm WK}(\varepsilon)
+ 2 \sum_{M=1}^{\infty} (-1)^M \bigg[ \bigg( \frac{3}{4 \pi^2 M^2}
\bigg(\frac{\varepsilon}{\hbar \omega}\bigg)^2 - \frac{3}{8 \pi^4 M^4} - 
\frac{1}{16 \pi^2 M^2} \bigg) \cos \bigg(2 \pi M \frac{\varepsilon}{\hbar
\omega} \bigg) \nonumber \\
&+& \bigg( \frac{1}{2 \pi M} \bigg(\frac{\varepsilon}{\hbar \omega}\bigg)^3 - 
\frac{3}{4 \pi^3M^3}\frac{\varepsilon}{\hbar \omega} -
\frac{1}{8 \pi M} \frac{\varepsilon}{\hbar \omega}\bigg) \sin \bigg( 2 \pi M
\frac{\varepsilon}{\hbar \omega} \bigg) \bigg] \hbar \omega.
\label{eq24} \end{eqnarray}
Therefore, in this simple model the separation of the total energy
in a smooth (liquid drop like) part $\tilde{E}$ and a shell correction
part $\delta E$, like in Eq.(\ref{eq1}), is obtained analytically.

The upper panel of Fig.1 displays the accumulated level density
$N(\varepsilon)$ for a set of fermions in a fixed {\it spherical} HO
potential calculated semiclassically and quantally, as a function of
the Fermi energy $\varepsilon$ divided by $\hbar \omega$. The quantal
result exhibits discontinuities at each major shell ($N= 2$, 8, 20,
40, 70, and 112 in the figure) and is represented by a staircase
function formed by horizontal and vertical lines which fluctuate
around the smooth value of $N(\varepsilon)$. The latter is provided by
the WK value given by Eq.(\ref{eq20}) and is represented by the solid
curve of the upper panel of Fig.1.
In the same panel we display the oscillatory part of $N(\varepsilon)$
(dashed curve), i.e., the quantal minus the semiclassical values,
which contains the fluctuations due to the shell effects. One sees
that the quantal part of the accumulated level density oscillates
around zero.

The lower panel of Fig.1 displays the quantal and semiclassical WK
values of the total energy $E(\varepsilon)/\hbar \omega$ for the
spherical HO potential by the staircase and solid curves,
respectively. In the same lower panel, the shell energy, i.e. the
difference between the quantal and semiclassical values, is
represented by the dashed line. Again it can be seen that the shell
energy fluctuates around zero and that the semiclasical WK estimate of
$E(\varepsilon)/\hbar \omega$ averages the quantal values. As it has
been pointed out in the Introduction, the WK approach to
$E(\varepsilon)$ including $\hbar^4$ corrections coincides with the
Strutinsky average in the HO potential \cite{BP}.

We have performed the same kind of analysis for the more realistic
potential well of Woods-Saxon (WS) type used in Ref. \cite{JBB}:
$V(r)=V_0/[1+\exp{(\frac{r-R}{a})}]$ with the values $V_0=-44$ MeV,
$a=0.67$ fm and $R=1.27 A^{1/3}$ fm. We have computed quantally and
semiclassically (with pure TF and with WK up to $\hbar^2$ order) the
accumulated level density and energy of neutrons (spin degeneracy is
assumed) in the above WS potential with a size corresponding to a 
nucleus of $A=208$ nucleons. The calculated
$N(\varepsilon)$ and $E(\varepsilon)$ are displayed as a function of
the Fermi energy $\varepsilon$ in the upper and lower panels of Fig.2,
respectively. Again the staircase and solid curves correspond to the
quantal and WK results, respectively, whereas the dashed lines are now
the pure TF values. As in the case of the HO potential, the WK
estimate of the smooth parts of $N(\varepsilon)$ and $E(\varepsilon)$
passes well through the corresponding staircase functions and averages
the quantal accumulated level density and energy.

For the WS potential the equivalence between the semiclassical WK
expansion and the Strutinsky average cannot be established
analytically. It has been checked numerically that both methods, with
high accuracy, give the same value for the energy, at least in the
case where the chemical potential is sufficiently negative
\cite{RS,CVDSE,CPVGB,JBB}. However, the situation may be different
when the Fermi energy is close to zero. In this case, the
semiclassical WK and Strutinsky level densities start to deviate from
one another when $\varepsilon$ approaches zero. The WK level density,
which includes $\hbar^2$ corrections shows a ${\varepsilon}^{-1/2}$
divergency at $\varepsilon=0$ for a finite potential as the WS one,
whereas the Strutinsky averaged level density only has a strongly
pronounced, but finite, maximum \cite{S}. In Refs. \cite{NWD,VKLNSW}
it was concluded that the divergency of the WK level density for
$\varepsilon \to 0$ is unphysical and preference should be given to
the Strutinsky smoothed level density. However, we would like to
recall that WK quantities have to be understood in the sense of
distributions \cite{RS,KCM}. Therefore, a diverging WK level density
should not be taken literally and only used under integrals. For
example, in the upper and lower panels of Fig.2 one sees that the
accumulated semiclassical level density $N(\varepsilon)$ and the total
energy $E(\varepsilon)$ are well behaved and accurately average the
corresponding quantal values even for $\varepsilon \to 0$. The TF
accumulated level density and energy show similar tendencies to those
exhibited by the WK results. However, the TF average of the quantal
values is less good than the one obtained at the WK level. This fact
demonstrates the importance of the $\hbar$-corrections in the Wigner
function (\ref{eq15}) to obtain the correct average of the quantal 
results. 

Usually the various quantities like energy, kinetic energy, etc. for a
system containing a fixed number of particles $N$ are not displayed as
a function of the chemical potential $\mu$ [given by $N=N(\mu)$], but
rather as functions of the particle number. For example, having the
energy $E(\mu)$ and the accumulated level density $N(\mu)$ as
functions of the chemical potential $\mu$ we can consider the
inversion $\mu=\mu(N)$ and then obtain the energy as a function of the
particle number $N$, i.e $E=E(N)$. The $N$ dependence can be studied
for a fixed external potential. More realistically the potential well
will change with the number of particles, as e.g. the HO potential
with $\hbar \omega \simeq 41 A^{-1/3}$ MeV or the WS potential of Ref.
\cite{BB}. Below we will consider both cases: the most simple case of
a fixed potential well and the case where the potential changes with
the particle number.

If the potential well has degenerate levels, the inversion
$\mu=\mu(N)$ is not unambiguous in the quantal case, because the
chemical potential is the same for various values of the particle
number $N$. This is for example the case for the spherical HO. To get
around this problem one can consider the spherical HO as the limit of
a triaxially deformed HO in the limit of zero deformation. In the
triaxial case each level has only spin-isospin degeneracy. However,
for the purpose of our reasoning we here can disregard spin and
isospin. Then in the infinitesimal triaxially deformed HO all levels
can be occupied by only ``one nucleon''. In the case of sphericity a
major shell with HO principal quantum number $n$ has a degeneracy
$D(n)$ and the functions $N(\mu)$ and $E(\mu)$ are sharp staircase
functions, whereas for very small triaxial deformation the vertical
jumps become slightly tilted and resolved in $D(n)$ minuscule
staircases. In that case one then always has a definite number of
particles for definite values of $\mu$ and perfectly can find
$\mu=\mu(N)$ unambigously. Therefore also $E(N)$ is well defined. In
the limit of zero deformation this leads to the uniform filling
prescription of a degenerate shell at sphericity.

With these preliminaries in mind, we show in the upper and lower
panels of Fig.3 the energy per particle as a function of the particle
number for (i) a strongly deformed HO with frequencies
\begin{equation}
\omega_x= \sigma^{-1/3} \delta^{-1/2} \omega_0 , \qquad
\omega_y= \sigma^{-1/3} \delta^{1/2} \omega_0 ,  \qquad
\omega_z= \sigma^{2/3} \omega_0 ,
\label{eq25} \end{equation}
taking the values $\omega_x=0.460 \, \omega_0$, $\omega_y= 1.111 \,
\omega_0$, and $\omega_z= 1.954 \, \omega_0$, and (ii) a spherical HO
in the sense explained above. The
HO depends as usual on particle number through $\hbar \omega_0 = 41
A^{-1/3}$ MeV and deformation is such that volume is conserved
($\omega_x \omega_y \omega_z= \omega_0^3$).
In the upper panel of Fig.3 we see that in the deformed
potential \cite{BB} the
quantal values (dots) practically coincide with the WK values
(solid line) and in any case WK perfectly averages the quantal
values. On the other hand, in the spherical case there is a surprise
in the sense that the WK-values do not pass, as a function of the
particle number, through the average of the quantal values: there are
much more values above the WK-line than below and also the deviations
above the WK-line are stronger than below. This means that WK
overbinds with respect to the true average except at magicity.

In the light of the fact that for the separate curves $E(\varepsilon)$
and $N(\varepsilon)$ (see Fig.1) the semiclassical values perfectly
average the quantal ones also in the spherical case, the
global overbinding
of WK as a function of the particle number may appear puzzling. The effect is, 
however, known \cite{Leb}. One can indeed show that an average over
$\varepsilon$ (or $\mu$) of the fluctuating part in (\ref{eq24})
yields zero, whereas when expressed as a function of
$N$ the fluctuating part shows a non-vanishing average, i.e. $\langle
\, \delta E(\mu) \, \rangle_{\mu} = 0$ but $\langle \, \delta E(N)
\, \rangle_N \ne 0$, where the brackets
$\langle \, \dots \, \rangle_{\mu,N}$ indicate averages over $\mu$ or
$N$, respectively. This feature 
can also be understood schematically from a different aspect in the 
following way.  
Suppose we consider a HO potential of fixed size with very small triaxial
deformation, i.e. we consider the uniform filling prescription at
sphericity. In a given shell the total quantal energy increases
linearly with the number of nucleons in the shell. On the other hand,
on average the total energy, according to Eqs. (\ref{eq20}) and
(\ref{eq21}), increases at the TF level as $E_{tot} \propto N^{4/3}$.
This situation is detailed in Table~1 for the $n$=4 shell of a
spherical HO potential of fixed size, which contains the $1g$, $2d$
and $3s$ levels. There we display the quantal and semiclassical (TF
and WK including $\hbar^2$ corrections) chemical potentials and
energies obtained in filling uniformly the shell assuming spin
degeneracy (the values are expressed in units of $\hbar \omega$). The
semiclassical chemical potentials are obtained inverting
Eq.(\ref{eq20}) to find the corresponding value of $\mu$ which in turn
is used in (\ref{eq21}) to calculate the semiclassical energies. The
quantal chemical potential in each spherical shell of the HO potential
is given by $\mu/\hbar \omega =n+3/2$. The number of particles and
energy in the $n$=4 shell are given by
\begin{equation}
N= \sum_{n=0}^{3} D(n) +2m = 40 + 2m
\label{eq24a} \end{equation}
and
\begin{equation}
\frac{E}{\hbar \omega} = \sum_{n=0}^{3} D(n)\bigg(n + \frac{3}{2}\bigg) 
+2m \bigg(4 + 
\frac{3}{2}\bigg) =  150 + 2m \bigg(4 + \frac{3}{2}\bigg) ,
\label{eq24b} \end{equation}
where $D(n)=(n+1)(n+2)$ is the degeneracy of a shell including spin
and $m=1,2,3, \ldots$ is the number of pairs (spin up and spin down)
added to fill up the $n$=4 shell.
From Table~1 we see that the TF energies always overbind the quantal
values and the same is true for the WK ones, except close to magicity
where the $n$=4 shell is empty or completely full. 

The shell correction, i.e. the difference between quantal and semiclassical
energies, is displayed in Fig.4 as a function of the number of
particles in the shell. From this figure it is clear that the TF
approach is far from averaging the quantal values and that in the WK
case the average is much improved. If the $\hbar^4$ corrections are
added, in this case of a fixed HO potential, the energy is shifted
down by a constant amount of 17/960 (in $\hbar \omega$ units)
according to Eq.(\ref{eq21}), but it cannot be distinguished from the
$\hbar^2$-corrected result on the scale of the figure. Therefore, the
$\hbar^4$ corrections are very small as compared with the
$\hbar^2$-ones demonstrating again the rapid convergence of the WK
series. Thus, the situation for a fixed HO potential is similar to
that found in the more general case of a size dependent HO potential
as it can be seen comparing the lower panel of Fig.3 with Fig.4.

The lack of averaging in the energies found in the spherical potential
is due to the large degeneracy of the HO shells. If each shell is
broken into $D(n)$ small pieces, as in the case of strong triaxiality,
they are bound to stay close to the average which varies, as already
mentioned, as $\sim N^{4/3}$. In the case of sphericity this $N^{4/3}$
behaviour is, as demonstrated in Table 1, 
quantally replaced by straight line segments, each
segment corresponding to a major shell. Two segments join at a magic
number with a characteristic overbinding which is relatively small. In
between two magic points the quantal straight line passes most of the
time {\it above} the concave semiclassical curve. This scenario can
further be clarified by the following investigation. The fact that for
strong triaxiality quantal and semiclassical calculations almost agree
can be understood because in that case there do not exist degeneracies
(besides some special cases where the axis ratios are formed by
rational numbers \cite{BJ}). Therefore, the quantal level density is
also practically smooth, and it almost coincides with the
semiclassical result.

In Fig.5 we show this, displaying the energy per particle as a
function of triaxial deformation for a HO well. To have a single
deformation parameter $d$ for the representation, in this figure we
have chosen the frequencies of the deformed HO according to
\begin{equation}
\sigma=1 + d \sqrt{3} ,  \qquad
\delta=1 + \vert d \vert \sqrt{2}
\label{eq25a} \end{equation}
in Eq.~(\ref{eq25}). We see that for a mid-shell configuration
(spin degeneracy)
of $N=92$ fermions, the semiclassical and quantal values practically
agree, up to very small fluctuations, down to quite low deformations.
The quantal energy suddenly raises when approaching
sphericity. For real nuclei this means that binding energy is lost at 
sphericity. The only slight exception to this scenario is for
deformation $\sim$ 0.6 where the frequency ratio is close to
$\omega_x: \omega_y: \omega_z \sim 1:2:3$. We, therefore, see that in
forcing open shell nuclei to be spherical one loses a lot of binding energy.
As a matter of fact, this loss of binding energy starts immediately
off magic numbers and increases towards mid shell fillings. This explains
why the semiclassical curve mostly overbinds as a function of the
shell filling. In the general case with mass number dependent
potentials, these considerations are slightly more complicated but the
reasoning which leads to the underbinding of quantal results for
energies per particle keeping open shell nuclei spherical is
essentially the same. This is one of the explanations of the fact that 
the shell corrections as a function of the particle number, do not oscillate
around zero but show a finite average value. Above we already have 
mentioned that this can also be seen from the fact that as function 
of the particle number the fluctuating part in Eq.(\ref{eq24}) does not
average to zero \cite{Leb}. Below we will find that the same situation
prevails in the case of self consistent mean fields. We also would
like to mention that similar features as those discussed above in
connection with deformation and degeneracy of the single-particle
levels have been found by Pomorski \cite{pomorski04} using the
Strutinsky smearing method applied in energy space and in particle
number space.

The above considerations only apply to spherical nuclei. In reality
the force which holds semimagic nuclei, e.g. tin isotopes, spherical
is the magicity of the protons which resists to deformation. For
deformed nuclei the situation is different and needs separate
investigation. In the lower panel of Fig.~3 we show what happens when
the shape of the potential is free and the energy minimised for each
particle number with respect to deformation. The absolute
minimum of the quantal calculation is obtained allowing triaxial
deformation in Eq.(\ref{eq25}).  Now the arches
are strongly flattened and in between magic numbers the energies per
particle lie practically on the semiclassical curve.
Magic nuclei appear as exceptional points and
a particle number average will be close to the semiclassical result.
Notice again that we are now comparing absolute minima both in the
semiclassical case (where they occur exclusively at sphericity) and in
the quantal one (where they are deformed, besides
around magicity). From this point of view, the close agreement between
quantal and semiclassical results is very satisfying and it likely is
a generic feature valid also for other types of mean field potentials.
For real nuclei, deformed as well as spherical situations can happen.
If both proton and neutron numbers correspond to open shell
situations, nuclei are in their majority deformed, whereas if either
the proton or neutron number is magic, nuclei usually are spherical,
as it happens for instance for the chain of Sn isotopes. Because of
the numerical complexity of the deformed case, we only will
concentrate on spherical nuclei in the remainder of the paper.
More detailed investigations of the deformed situation will be
presented elsewhere \cite{Leboeuf06}.

The fact that the semiclassical results are not going through the
average of the quantal results as a function of $A$ is somewhat
annoying from the practical point of view, since we cannot judge
whether the semiclassical results are converged to the right value or
not. In the case of external potentials the answer to this question is
easy to find: we take a fixed potential and look at the WK results as
a function of the chemical potential $\mu$. We know that in this case
the semiclassical results should pass through the average of the
quantal ones (see e.g. Fig.2).

\section{The self consistent potential case}

\subsection{Finite nuclei}

For spherical nuclei described self consistently through an effective
interaction, the scenario for the energy per particle as a function of
the mass number stays qualitatively the same as for the external
potential case. Again, the typical arch structure with the values at
magicity barely undershooting the semiclassical line (see lower panel
of Fig.3) appears. In the upper panel of Fig.6 we present
self consistent calculations of
the shell energy per particle, which is defined as $E_{\rm shell}/A =
(E_{\rm HF} - E_{\rm semicl})/A$, as a function of the mass number,
for the TF, VWK2, and ETF4 semiclassical approaches using the T6 force
\cite{T6}. This Skyrme interaction has an effective nucleon mass $m^*$
equal to the bare one $m$. In the calculations shown in Fig.6, the
Coulomb repulsion among protons and the spin-orbit force have been
switched off and only hypothetical spherical symmetric nuclei with
$N=Z$ are considered. Later we will study realistic nuclei, but for the
moment we want to avoid that these more subtle effects contaminate the
comparison of the semiclassical results with the quantal ones.

The same type of calculation is presented in the lower panel of Fig.6,
now performed
using the very different Skyrme force SV \cite{Sk5} that has no
density-dependent part ($t_3=0$) and for which the effective mass is
$m^*/m$=0.38 in nuclear matter. In the case of the VWK2 calculation we
encounter the same pattern as with the T6 force. However, the
predictions of the TF calculation are at variance with the case of the
T6 interaction: for T6 they overbind, whereas for SV they underbind.
This change in the behaviour of the TF solution is largely due to the
different  values of the effective masses of the two forces, and we
have documented this fact already in earlier publications
\cite{CVDSE,CVBML}. Very satisfactorily, the deviation of the VWK2
results from the HF ones only depends very little on the particular
properties of the effective interaction.

We have to remark the important feature that, as can be seen in Fig.6,
the shell energies per particle calculated with the VWK2 and
ETF4 approaches are very close. We have tested a whole series of
Skyrme forces and did not find exceptions to this fact. As we will see
below, the inclusion of Coulomb and spin-orbit forces does not change
this rule essentially. Thus, with the VWK2 calculation one is able to
obtain energies of an equivalent quality to the by far more
complicated ETF4 approach which requires sophisticated techniques for
the self consistent numerical solution of the variational equations.

Several additional comments should, however, still be made. Above we
argued that ETF is inconsistent in sorting out the powers in $\hbar$
and that consequently it converges less well than VWK\@.
We also gave arguments backed by explicit examples that VWK should
converge faster than ETF and that the $\hbar^4$-contribution to VWK
should practically be negligible. This was, however, for an external
potential case. Perhaps the external potentials are particularly
difficult cases for ETF (for instance one has to solve a non-linear
differential equation even in the external potential case, whereas the
WK expression can be used as is) and its convergence properties are
better in the completely self consistent case. Thus, it could be that
both the VWK2 (besides a small VWK4 correction as in the external
potential case) and the ETF4 results are converged to the same and
definite semiclassical value for nuclear binding energies. However, at
this point this is a speculation. It may be that VWK2 and ETF4
coincide without both really having reached complete convergence to
the actual semiclassical average. For instance, it cannot be excluded
that VWK4 could, in the self consistent case, yield a contribution
which is sensitively more important than in the external potential
case. This remark should be kept in mind when we discuss the results
more closely below.

The uncertainty of the situation also comes from the fact that, as
discussed above, we do not have a precise criterion in the
self consistent case what the semiclassical binding energies as a
function of the particle number should be, besides that they should
coincide with a Strutinsky self consistent calculation. The latter is,
however, also slightly uncertain, because the plateau condition is
difficult to satisfy \cite{BB,Pom} and up to
date only a few self consistent Strutinsky calculations with Skyrme
forces exist in the literature \cite{BQ,BQ1}. However, in any case the
general  agreement of VWK2 with ETF4 is quite remarkable and adds more
confidence to the semiclassical results, in spite of the fact that 
problems are not all resolved as we will further discuss below.

In order to give additional backing to what is just outlined with more
studies, we now show on the upper panels of Figs.7 and 8 the energy
and the shell correction per nucleon, respectively, in a realistic
case for nuclei along the valley of stability calculated with the
SkM$^{*}$ interaction including the Coulomb and spin-orbit forces.
Most of the $\beta$-stable nuclei displayed in Figs.7 and 8 were
also used in Ref. \cite{Pom}, and they are spherical according to the
finite range droplet model (FRDM) \cite{FRDM1}. We have also
considered some other additional nuclei which, according to the FRDM,
are also $\beta$-stable \cite{FRDM1} and spherical except for a few of
them \cite{FRDM2}. Also in this realistic case we can observe the very
close agreement between the VWK2 and ETF4 methods. We also display the
results of the ETF2 calculation and notice that it is much less
converged. The VWK2 approach to the relativistic mean field theory has
also been worked out \cite{ECV} and we present the results for the
same sample of nuclei along the valley of stability using the
realistic, accurately calibrated
parameter set NL3 \cite{NL3} on the lower panels of Figs.7
and 8. In the relativistic case the ETF4 corrections have not been
elaborated and we do not have the corresponding results available. We
have found the relativistic ETF2 results to be, as in the
non-relativistic case, not so well converged as VWK2 and we do not
show them.

In Fig.8 we quite clearly see a deficiency of the semiclassical
energies, already present in the preceding figures: there is too much
binding, even keeping in mind that the semiclassical binding energies for 
spherical nuclei have a natural tendency to be stronger than the average of 
the quantal values as discussed in section III. This drawback is particularly 
pronounced in the case of Skyrme forces (we checked this with multiple
Skyrme interactions). Also in the relativistic case the semiclassical
results give too much binding, in spite of the fact that the doubly
magic nuclei are (slightly) more bound quantally than semiclassically,
what is not the case for Skyrme forces, see Table 2. This situation
clearly is unphysical. We will comment further on this in the next
subsection. Let us point out that in our prescription the shell
correction has been taken as $E_{\rm shell} = E_{\rm HF} - E_{\rm
semicl}$, which in principle is different from the often employed
prescription where one takes $E_{\rm shell} = \sum \varepsilon_i -
\sum \varepsilon_i \tilde {n}_i$ as the difference of the sum of the
quantal single-particle energies and the Strutinsky averaged sum.
However, due to the Strutinsky energy theorem \cite{RS} the
predictions of both procedures should agree if the considered
semiclassical approaches reproduce well the Strutinsky averaged value.

Concluding this subsection, we can say that the semiclassical limit of
the energy per particle of finite nuclei based on Skyrme or
relativistic mean field theories has been established on the VWK2
level. A quite intriguing coincidence between the VWK2 and ETF4
methods has been found. The significance of this fact is not entirely
clear and will be discussed further in the next section. The ETF4
method exists since long whereas VWK2 is new. Apart from the discussed
conceptual differences in the rigorous power counting in the $\hbar$
expansion, the VWK2 method, see Eq.(\ref{eq12}), has the advantage
that the convergence is faster and that the final formulas for the
calculation $E/A$ are very simple (only the solution of the
zeroth-order TF variational equation is needed!). The overbinding of
TF and ETF calculations of nuclei has been recognized in many studies
since years ago (see e.g. Refs.\ 
\cite{BGH,CPVGB,PS,Li,BB,CVBS,CV,Kumar,BCKT}), 
and it is also very much present in atomic physics calculations
\cite{March}. We have shown that the problem persists even in the more
refined ETF4 and VWK2 appoaches, and we will turn to it in more detail now.

\subsection{The overbinding problem}

In the last section we have seen that the scenario of the arch
structure in the energy per particle remains in the self consistent
case qualitatively the same as in the external potential case.
However, we remarked a deficiency in the self consistent case which
becomes apparent when having a close look at the realistic cases
presented in Figs.7 and 8. In Table 2 we present the quantal and VWK2
energies for some magic nuclei calculated with the SkM$^*$ force and
with the NL3 parameter set of the relativistic theory. We see, that in
particular Skyrme forces overbind semiclassically, as it is also seen
in the upper panels of Figs.7 and 8. The fact that even doubly magic
nuclei like $^{40}$Ca are more bound semiclassically than quantally is
clearly incorrect, even though we should be aware of the fact that we
are dealing with small differences of large numbers. In any case,
taking self consistent Strutinsky calculations as reference
\cite{BQ,BQ1},  $^{40}$Ca and $^{208}$Pb are less bound, when
averaged, than quantally. This failure of the semiclassical approach
is disappointing with respect to the external potential case where, as
a function of energy (or chemical potential), we are used to the fact
that the VWK2 method gives extremely precise average quantities, like
level densities, energies, etc., as has been demonstrated in the past
with many examples \cite{RS,BP,SV,CVDSE} (see also Figs.1 and 2 of
this paper). 

Let us try to find some reason for this deficiency and eventually a
cure. What is different between the external and self consistent
potential cases? The only convincing difference we can imagine is the
fact that in the external potential case the density is a functional
of an external {\it fixed} potential $V$, i.e. $\rho=\rho[V]$ which is
expanded in powers of $\hbar$, that is in gradients of $V$. In the
self consistent case the potential itself is a functional of the
density and therefore also the potential has then to be expanded in a
power series of $\hbar$ (we did not explicitly proceed in this way,
but implicitly that is what it amounts to in the VWK method). This
double $\hbar$-expansion is very likely one of the reasons for the
deterioration of the results with respect to the external potential
case. The WK expansion of the density matrix is a local expansion in
terms of distributions functions what probably does too much harm to
the self consistent potential. Some global features should be kept,
even for the average potential. For example, if we were given a self
consistently averaged Strutinsky HF potential, we would believe from
our past experience that when it is taken as an external potential in
the evaluation of the semiclassical WK-HF energy this should give very
precisely the true Strutinsky averaged value of $E/A$. Of course, this
would be an extremely laborious detour. One can think of employing an
approximate substitute of the Strutinsky potential. A possibility is
to take the self consistent potential evaluated in the ETF2 approach,
instead. Indeed, as we mentioned earlier, in the ETF2 approach the
density contains powers in $\hbar$ which are partially resummed to all
orders in the self consistent calculation \cite{SV}. Therefore, the
corresponding single-particle potential, which is a well behaved smooth 
average potential (see Fig.12), also contains some global properties.

We will apply this strategy to obtain another semiclassical estimate
of the HF energy. To this end we first run a self-consistent ETF2
calculation with the T6 force. Next, we take the computed ETF2 mean
field potential, including its spin-orbit and Coulomb contributions,
as if it were an {\it external}\/ potential and with it we perform a
WK2 calculation to obtain the Skyrme energy, by using the WK
expressions for particle and kinetic energy densities including
$\hbar^2$ corrections \cite{RS}. In this procedure, which clearly
differs from the VWK method, some divergences arise in the evaluation
of some $\hbar^2$ contributions (see Appendix 1 for the treatment of
the divergence of the term $( \nabla \rho_{\rm WK} )^2$ in the Skyrme
energy density). To circumvent this technical difficulty, a finite
temperature WK calculation is performed \cite{BGH,VG} which is
extrapolated to $T=0$. The details of this method to estimate the
semiclassical HF energy are described in Appendix 2. The results
obtained for the shell correction per particle calculated for
$\beta$-stable nuclei are shown in Fig.9 (curve labelled by
``temperature extrapolation''). We see that there is a substantial
improvement over the VWK2 and ETF4 results. The cure is not 100\%
though and there remains the fact that $^{40}$Ca is slightly more
bound semiclassically than quantally. However, globally, the average
$A$-dependence of $E/A$ is now quite acceptable in particular towards
the heavier nuclei. For example the shell correction for $^{208}$Pb is
now $\sim 20$ MeV, well in line with the value reported from
Strutinsky calculations with the Gogny D1S and RMF NL3 effective
nuclear interactions \cite{Pom}. We agree that the procedure is ad hoc; 
however, it helps to shed some light on the situation.

It is interesting to note that the WK-HF results in Fig.9 obtained
with the ETF2 potential can almost perfectly be reproduced within the
VWK2 method with a fudge factor on the VWK2 kinetic energy in the
following way: in Eq.(\ref{eq12}) we replace the factor 1/36 in front
of the Weizs\"acker term by 1.26/36 in order to increase the kinetic
energy, i.e. to decrease the binding. The same result can be obtained
with ETF2 by replacing 1/36 by 1.8/36 (modifications of the value of
the coefficient of the Weizs\"acker term have been studied in the
literature since a long time ago, see e.g. Ref.\cite{KT}). We show
these results for the T6 force in the upper panel of Fig.10 and find
that all these different prescriptions practically lead to shell
correction values on top of one another. The reason for this very
close agreement, found using different prescriptions to estimate the
shell corrections, is at present unknown, but it is a surprising and
interesting feature.

Still we would like to deepen somewhat the discussion of the situation
and of the possible reasons for the failure of VWK2
(and ETF4) to correctly reproduce the average.
As mentioned above, the implicit expansion of the average mean field   
in powers of $\hbar$ may be the main direct reason. However, the very 
short range character of the nuclear force may reinforce the problem
at least in the non relativistic case. 
For Skyrme forces the zero range character entails an unphysical shape of 
the self consistent TF potential. This can best be studied in half infinite 
nuclear matter where the self consistent TF density can be obtained in an 
analytic way by quadratures in the case of Skyrme forces \cite{CVDSE,VCDS}. 
The TF density $\rho_0(z)$ and the corresponding single-particle potential for
a Skyrme force with $m^* = m$ 
\begin{equation}
V(z) = 2 a_0 \rho_0(z) +
(2 + \alpha) a_3 {\rho_0(z)}^{1 + \alpha} 
- 2 a_{12} \, \rho''_0(z),
\label{eq35} \end{equation}
are displayed in Fig.11.
 The TF density close to the turning point, chosen at 
$z=0$, behaves like $\lim_{z \to 0} \rho_0 \approx z^2$ and the 
single-particle
potential $V(z)$ reaches the classical turning point at $z=0$ with zero slope.
This feature is not very well seen in Fig.11 because it turns
out that the bending into the horizontal tangent only happens extremly close 
to the classical turning point.
On the other hand, such 
pathological behaviour is absent in the relativistic mean field approach, 
since the forces are of finite range. In this case the TF potential has a
WS like shape and is continuous in whole space. This is shown in Fig.12 where
we display the TF neutron self consistent potential for $^{208}$Pb 
obtained with the NL3 parameter set. We see that this potential has a very 
acceptable shape, not much different from the usual phenomenological WS potentials 
with about a 2 fm wide fall off width. Also the derivative of this potential 
is in no way anywhere more pronounced than the one corresponding to 
phenomenological potentials (see Fig.12). From this fact we understand
that,
with respect to the non-relativistic case of Skyrme forces, the semiclassical 
results are considerably better in the relativistic case (see Table 2 and
Figs.8 and 9). At least practically all doubly magic nuclei are more bound 
quantally than semiclassically. However, for example for $^{208}$Pb the 
shell energy turns out to be $E_{shell} \approx 6$ MeV, a value which
is about a factor 3 times too small with regard to commonly accepted
values \cite{Pom,BQ,BQ1}.
Again this problem may be attributed to this double expansion in
gradients of the mean field potential and the potential itself and, as already 
mentioned, it cannot be excluded that the $\hbar$-expansion converges in the 
self consistent case more slowly than in the the external potential case. At 
any rate, contrary to the situation with the Skyrme forces, in the 
relativistic case, as already stated, the TF mean field potentials are 
perfectly smooth and well behaved and can, therefore, not be incriminated. 
At the present moment it is unclear how to remedy this situation, 
other than by prescriptions such as the ones presented above. However, even 
this  must still be refined in order to become entirely realistic.

A last comment may be in order at this point. We remark in Fig.10
that  for $^{208}$Pb the shell correction predicted by the ad-hoc
methods is quite acceptable. However, there is a continuous
deterioration of the situation towards lighter nuclei. Such a
deterioration can in fact also be seen in Fig.8 for the SkM$^*$ force
and in Fig.9 for the T6 force, whereas in the relativistic case (lower
panel of Fig.8) the predictions for light nuclei are more robust. For
instance, both SkM* and T6 yield (wrongly) a positive shell correction
energy of about 12 MeV for $^{40}$Ca in VWK2, while NL3 at least
predicts a negative value of $-2$ MeV for this nucleus (see Table 2).
One possible explanation for this different behaviour could be the
different  treatment of the spin-orbit potential in both cases. In the
non-relativistic case one should realize that the spin-orbit is not
only expanded in powers of $\hbar$ but in addition one assumes the
smallness of the coupling constant and only the lowest order term is
taken into account. In reality the spin-orbit is a matrix problem as
recognized by Frisk \cite{Frisk} and then no expansion in the coupling
constant is needed. To our knowledge, the validity in the expansion of
the coupling constant has never been checked. On the other hand in the
relativistic case such an expansion is absent, and the coupling of the
spin-orbit is treated at all the orders even in the semiclassical
approach  \cite{CVBS}. It is an open hypothesis whether this
difference  can explain the different behaviour in the upper and lower
panels of Fig.8. The spin-orbit potential yields a surface
contribution and this could point to the fact why in one case things
deteriorate towards lower mass nuclei whereas in the other not. More
studies on this issue are needed.

On the above grounds, it seemed interesting to us to also apply a
fudge factor to the kinetic energy in the relativistic case. With the
very small coefficient 1.025 we obtain the results shown in the lower
panel of Fig.10. Now the shell energy of $^{40}$Ca is 5.6 MeV and the
one of $^{208}$Pb is 15 MeV. Both results are compatible with
previously known values \cite{Pom,BQ,BQ1}. We therefore have now at
hand, at least in the relativistic case, an ad hoc procedure which
yields reasonable shell energies throughout the periodic table. This
very small correction needed in the relativistic case may hint to the
point that there the $\hbar^4$-corrections could cure the overbinding
problem with no need of a fudge factor. However, for $^{208}$Pb still
$\sim 10$ MeV overbinding in the total energy occurs semiclassically
whereas for an external potential the $\hbar^4$-corrections to the
energy are typically $\sim 1$ MeV only \cite{RS}.

The conclusion of our study therefore is that the semiclassical method
based on the asymptotic expansion of the Wigner-Kirkwood type is in
the self consistent case more fragile, i.e. inaccurate, than in the
external potential case. This failure in the case of finite nuclei is in
agreement with our earlier studies on the surface energies
\cite{SV,CVDSE,ECV} which also turn out to be much more accurate in the
external potential case than in the self consistent one. These findings are,
however, sensitively more pronounced in the case of Skyrme forces than in the
case of relativistic mean field theory.

\newpage

\section{Conclusions} 

In this paper we took up again the old problem of the Thomas-Fermi
approach to nuclei with incorporation of $\hbar$-corrections. As we have
pointed out in earlier work \cite{SV}, the well established ETF scheme,
lacking a correct sorting out of powers in $\hbar$, may show
unnecessary slow convergence properties. We therefore established a
rigorous order by order $\hbar$ expansion of the self consistent
nuclear mean field problem which we named variational-Wigner Kirkwood
(VWK) theory. We here apply it for the first time to finite nuclei in
realistic self consistent mean fields at order $\hbar^2$ (VWK2),
supposing that the $\hbar^4$-corrections be very small, similarly to
what is documented for the external potential case since several
decades \cite{RS}.

One essential finding of our investigation is that practically for all
Skyrme forces VWK2 yields binding energies per particle very close to
fourth order ETF functional theory (ETF4). This result is good and bad
at the same time. The good point is that the results from ETF4 can be
reproduced with the much simpler VWK2 approach and that the agreement
gives further credit to the correctness of the semiclassical values.
The bad side is that it is known since long
\cite{BGH,CPVGB,PS,Li,BB,CVBS,CV,Kumar,BCKT} that ETF even at order
$\hbar^4$ produces E/A values with too much binding yielding for
instance for doubly magic nuclei (e.g. $^{40}$Ca or $^{208}$Pb) values
which are lower in energy than the ones obtained from quantal
Hartree-Fock calculations. Evidently this overbinding problem is then
also present in VWK2.

We advanced several arguments in regard to the overbinding problem
such as the zero range character of the Skyrme forces, leading to
unphysical shapes of the self consistent Thomas-Fermi mean field
potential, and/or an insufficient treatment of the spin-orbit
potential. Those arguments are backed by the fact that in the
relativistic RMF, with finite range meson exchange potentials, the
situation is considerably better. Indeed in that case at least
practically all of the doubly magic nuclei are more bound quantally
than semiclassically. This could stem from the fact that there the
Thomas-Fermi mean field potential is perfectly smooth resembling very
much a realistic Woods-Saxon type of potential. Also the spin-orbit is
treated properly. However, the shell corrections, e.g. for $^{208}$Pb,
are with relativistic VWK2 (no ETF4 exists in that case) still roughly
a factor of three too small. This remaining failure could have as
origin that in the self consistent case, contrary to the external
potential case, the mean field is itself a functional of the density
and has to undergo an $\hbar$-expansion (this remark is also true in
the non-relativistic case). Evidently also missing
$\hbar^4$-corrections can be invoked. One has to keep in mind,
however, that for heavy nuclei $\hbar^4$-corrections are typically of
order 1 MeV in the external potential case, whereas even in RMF
semiclassical energies are about 10 MeV overbound. Still, a slower
convergence of the $\hbar$-expansion in the self consistent case is
not to be excluded.

Since $\hbar^4$-terms enormously complicate the theory and the
numerical treatment, we refrained from studying this here, and to gain
further insight we rather investigated whether the situation can be
improved by ad hoc prescriptions. We report on several possibilities,
where a fudge factor of 1.025 on the kinetic energy in the
relativistic case gives the most satisfying results. Indeed, we show
in the lower panel of Fig.10 the corresponding shell energies for
spherical nuclei as a function of mass number which we believe are
quite realistic throughout. Probably, if in the non-relativistic case
finite range forces were used, the situation also would improve there
and an approach like in Ref.\cite{Swia} {\it including}
$\hbar^2$-corrections could be undertaken. However, even the
Thomas-Fermi solution with the Gogny force shows pathologies, since it
still contains zero range pieces. We also should mention that for
simplicity our studies were done almost exclusively for spherical nuclei. 

A result on the side, obtained in the external potential
case, was that for spherical nuclei the semiclassical binding energies per
particle as a function of particle number do {\em not} pass through
the average of the quantal results. Rather the semiclassical curve
(see Fig.3) shows more binding than the average. This fact, though
not unknown \cite{Leb}, has not been mentioned much in the past. This
natural tendency of the semiclassical results to give in the spherical
case (for the deformed one, see Fig.3 and remarks towards the end of
Section III) more binding than the average should, however, not be
confused with the overbinding problem encountered in the self consistent
semiclassical approach as discussed above and in section IV.B. This is
an additional and erroneous binding contribution contained in the
present semiclassical expressions which brings in the self consistent case
semiclassical binding energies below the ones of doubly magic nuclei, in
contradiction with results from self consistent Strutinsky calculations
where this is not the case and which should be the gauge for semiclassical
results.

Our studies may have relevance not only for nuclear systems but also
 for atomic physics calculations and for all other inhomogeneous
and/or self-bound Fermi systems like $^3$He drops, trapped cold atoms,
metallic clusters, quantum dots, etc., where the application of
statistical Thomas-Fermi methods is particularly helpful and valuable.

\section{Acknowledgments}
The authors are indebted to R.K. Bhaduri, O. Bohigas, P. Leboeuf, and
W.J. Swiatecki for useful comments and informations. Especially we
thank P. Leboeuf for pointing out to us that the fluctuating part of
the energy in (\ref{eq24}) contains a non-vanishing contribution when
averaged over particle number and that this feature is worked out in
Ref. \cite{Leb}. This work has been partially supported by the
IN2P3-CICYT collaboration. Two of us (X.V. and M.C.) also acknowledge
financial support from Grants No.\ FIS2005-03142 from MEC (Spain) and
FEDER, and No.\ 2005SGR-00343 from Generalitat de Catalunya.

\newpage

\section{Appendix 1}

As is well known the WK $\hbar$-expansion of the particle density
involves divergent terms at the classical turning point of the kind
$\propto (\mu - V)^{-n/2}$ where $n$ is an {\it odd integer and
positive} number, $V(\vec{r})$ the external mean field potential, and
$\mu$ the chemical potential. It is well documented in the literature
\cite{RS,KCM} how to deal with this divergent part and to extract the
non divergent contribution of the corresponding integral. Actually, we
gave in Section 3 a recipe for solving this problem in the case of an
external potential. Another way consists in writing
\begin{equation}
(\mu - V)^{-n/2} \propto \bigg (\frac{\partial}{\partial \mu} \bigg)^{n'}
(\mu - V)^{-1/2} ,
\label{eqA1} \end{equation}
where $n'= (n-1)/2$ and $(\mu - V)^{-1/2}$ is an integrable
divergency, and the
differentation can then be done after the integration has been performed.

A case which is not so well studied is the one of $\propto (\mu -
V)^{-2n'}$, i.e. terms with integer powers of $\mu - V$ in the denominator.
Such terms arise for instance when some powers of the density have to
be integrated. We will show here how to extract the finite part of
such integrals. In particular in the Skyrme energy computed at the
WK-$\hbar^2$ level we have to find the integral of the following
expression (see Eq.(\ref{eq34}) of Appendix 2):
\begin{equation}
( \nabla \rho_{\rm WK} )^2 = I_0 + I_2
\label{eqA2} \end{equation}
with
\begin{equation}
I_0 = C^2 \bigg[ \frac{9}{4} (\mu - V) \big(\nabla V)^2 + \frac{3
\hbar^2} {16 m} \nabla\big(\Delta V\big) \nabla V \bigg]
\label{eqA3} \end{equation}
the non-diverging part and
\begin{equation}
I_2 = C^2 \frac{3 \hbar^2}{64 m} \bigg[ \frac{\nabla V \nabla
\big(\nabla V\big)^2 + 2 \Delta V  \big( \nabla V \big)^2}{\mu - V}
+ \frac{3}{2} \frac{\big(\nabla V \big)^2}{(\mu - V)^2} \bigg]
\label{eqA4} \end{equation}
the diverging one ($C$ is a constant). Without loss of generality we
can assume here the
spherically symmetric case and then we can write for the integral of 
(\ref{eqA4}) $i = i_1 + i_2$ with 
\begin{equation}
i_1 = B_1 \int^{r_c}_0 dr r^2 \frac{G_1(r)}{\mu - V}
\label{eqA5} \end{equation}
\begin{equation}
i_2 = B_2 \int^{r_c}_0 dr r^2 \frac{G_2(r)}{(\mu - V)^2}
\label{eqA6} \end{equation}
where $B_1, B_2$ and $G_1$, $G_2$ are well defined constants and functions, 
respectively.

A judicious and frequently used strategy to isolate the finite
contribution is to integrate by parts and disregard the diverging
integrated piece. We have
\begin{equation}
\frac{\partial}{\partial r} \frac{1}{\mu - V} = \frac{V'}{(\mu - V) ^2}
\, ; \qquad V'= \frac{\partial V}{\partial r} .
\label{eqA7} \end{equation}
Then we can write for the integral of $I_2$:
\begin{equation}
i_1 + i_2 = \int^{r_c}_0 dr \frac{B_1 {\tilde{G}}_1 - B_2 {{\tilde{G}}_2}'}
{\mu - V} = i ,
\label{eqA8} \end{equation}
where ${\tilde{G}}_1= r^2 G_1$ and ${\tilde{G}}_2= r^2 G_2/V'$.
We perform one further partial integration and write, with $x=r/r_c$,
$$\frac{\partial}{\partial x} \ln (1 - V/\mu ) =
- \frac{1}{\mu - V} \,
\frac{\partial V}{\partial x} .  $$
Again neglecting the integrated part one obtains
\begin{equation}
i = r_c \int^1_0 dx \frac{\partial F}{\partial x} \, \ln (1 - V/\mu) , 
\label{eqA9} \end{equation}
where
\begin{equation}
F(x) = \big[B_1 {\tilde{G}}_1 - B_2 {{\tilde{G}}_2}' \big]
\frac{1}{\partial V/\partial x} .
\label{eqA10} \end{equation}
For example, for a Woods-Saxon potential the integral in
Eq.(\ref{eqA9}) can
directly be performed numerically. In the case of a harmonic oscillator 
potential $V=a r^2$ the integral can be performed analytically:
\begin{equation}
i = \gamma \big[ \ln 2 - \frac{4}{3} \big],
\label{eqA11} \end{equation}
with 
$$\gamma = \frac{3}{16} \frac{\hbar^2}{m} A^2 a^2 r^3_c.$$

The above result is the value which is obtained looking up integral
tables. We now want to check the value of $i$ with a second
independent method. This can be done writing the integral $i$ at
finite temperature (see Appendix 2), where it is not divergent, and
evaluating it as a function of $T$. At the end an extrapolation to $T
=0$ is performed. This method is also well documented in the
literature \cite{BGH} and will be briefly discussed in Appendix 2 for
the sake of completeness. However, the extrapolation process needs
some care and usually the final number will only be precise to a
couple of percent. Fig.13 displays, as a function of temperature, the
value of $\int (\nabla \rho)^2 d \vec{r}$ obtained with a HO potential
(upper panel) for $A=40$ and a WS potential (lower panel) for $A=90$.
As it is discussed in Appendix 2, the linear behaviour of this
integral with $T^2$ breaks down about $T=1$ MeV (as shown by the
curves that bend upwards in Fig.13), and the extrapolation to $T=0$
MeV (dashed lines of Fig.13) is needed. We
find that the extrapolated values are 1.789 fm$^{-5}$ and 2.494
fm$^{-5}$ for the HO and WS potentials respectively. The
values calculated with (\ref{eqA9}) give 1.867 fm$^{-5}$ (HO) and
2.383 fm$^{-5}$ (WS), which correspond to relative differences of
4.2\% and 7.4\%, respectively. Such errors are to be expected in the
extrapolation procedure.

\section{Appendix 2}

In this Appendix we present the details of the method which has been
used to obtain the results displayed by the semiclassical
``temperature extrapolation'' curve in Fig.~9 of Sect.~IV.B. It is
based on a calculation of the WK-HF energy including $\hbar^2$
corrections that is built on top of a previously computed smooth mean
field potential. The smooth potential, including the spin-orbit
($\vec{W}$) and Coulomb ($V_{\rm Coul}$) contributions, is generated
in a self consistent ETF2 calculation with the Skyrme T6 interaction.
It is then used as input for the WK calculation, where it is treated
as an external potential. To circumvent the divergence problems (see
Appendix 1) in some $\hbar^2$ terms of the WK-HF energy functional, we
compute the energy at finite temperature and take its limit
(numerically) when $T \to 0$. This has been shown in the past
\cite{VG} to be a very efficient procedure to overcome the
divergences. 

The expression of the WK energy in the case of a Skyrme force with
$m^* = m$, including Coulomb and spin-orbit contributions, reads
\begin{eqnarray}
E_{{\rm WK}} &=& \int d \vec{r} \bigg\{ \frac{\hbar^2}{2 m_n} \bigg[ \frac{3}{5}
\big(\frac{3\pi^2}{2}\big)^{2/3} \rho_{{\rm WK},n}^{5/3} + \frac{1}{36} 
\frac{(\nabla
\rho_{{\rm WK},n})^2}{\rho_{{\rm WK},n}}+ \frac{1}{3} \Delta \rho_{{\rm WK},n} \bigg] \nonumber
\\ &+& \frac{\hbar^2}{2 m_p} \bigg[ \frac{3}{5}
\big(\frac{3\pi^2}{2}\big)^{2/3} \rho_{{\rm WK},p}^{5/3} +
\frac{1}{36}\frac{(\nabla \rho_{{\rm WK},p})^2}{\rho_{{\rm WK},p}}+ \frac{1}{3} \Delta
\rho_{{\rm WK},p} \bigg] \nonumber \\
&+& \frac{1}{2}t_0 \bigg[ \big( 1 + \frac{x_0}{2} \big) \rho_{{\rm WK}}^2 -
\big( x_0 + \frac{1}{2} \big) \big( \rho_{{\rm WK},n}^2 + \rho_{{\rm WK},p}^2 \big) \bigg]
\nonumber \\
&+& \frac{1}{12}t_3 \rho_{{\rm WK}}^{\alpha}
\bigg[ \big( 1 + \frac{x_3}{2} \big) \rho_{{\rm WK}}^2 -
\big( x_3 + \frac{1}{2} \big) \big( \rho_{{\rm WK},n}^2 + \rho_{{\rm WK},p}^2 \big) \bigg]
\nonumber \\
&+& \frac{1}{16} \bigg[ 3t_1 \big( 1 + \frac{x_1}{2} \big) - t_2 \big( 1 +
\frac{x_2}{2} \big) \bigg] \big( \nabla \rho_{{\rm WK}} \big)^2 \nonumber \\
&-& \frac{1}{16} \bigg[ 3t_1 \big( x_1 + \frac{1}{2} \big) - t_2 \big( x_2 +
\frac{1}{2} \big) \bigg] \bigg[ \big( \nabla \rho_{{\rm WK},n} \big)^2 +
\big( \nabla \rho_{{\rm WK},p} \big)^2 \bigg]  \nonumber \\
&+& \frac{1}{2} W_0 \bigg[ \vec{J}_{{\rm WK}} \cdot {\nabla \rho_{{\rm WK}}}
+\vec{J}_{{\rm WK},n} \cdot {\nabla \rho_{{\rm WK},n}}
+ \vec{J}_{{\rm WK},p} \cdot {\nabla \rho_{{\rm WK},p}} \bigg]
+ {\cal{H}}_{\rm Coul}
\bigg\}.
\label{eq34}  \end{eqnarray}
In this equation $\rho_{{\rm WK},q}$ ($q= n,p$) is the WK neutron or
proton density including $\hbar^2$ corrections \cite{RS}, and
$\rho_{{\rm WK}}=\rho_{{\rm WK},n}+\rho_{{\rm WK},p}$. The neutron or proton semiclassical
spin-current density up to $\hbar^2$ order is given by \cite{GV,BGH}
\begin{equation}
\vec{J}_{{\rm WK},q}= - \frac{2m_q}{\hbar^2} \rho^{0}_{{\rm WK},q} 
\vec{W}_q ,
\label{eq34a} \end{equation}
where $\rho^{0}_{{\rm WK},q}$ is the TF ($\hbar^0$ part) of the WK neutron or
proton densities, $\vec{W}_q$ is the (external ETF2) spin-orbit
potential,
and $\vec{J}_{{\rm WK}}=\vec{J}_{{\rm WK},n} + \vec{J}_{{\rm WK},p}$. The
semiclassical Coulomb energy density appearing in Eq.(\ref{eq34}) is
computed as
 \begin{equation}
{\cal{H}}_{\rm Coul}  = \frac{1}{2} e^2 \rho_{{\rm WK},p}
V_{\rm Coul} 
- \frac{3}{4}\bigg(\frac{3}{\pi}\bigg)^{1/3} 
e^2 \rho_{{\rm WK},p}^{4/3} \,, 
\label{eq34b} \end{equation}
where $V_{\rm Coul}$ is the Coulomb potential provided by the 
self consistent ETF2 calculation.

At a finite temperature the relevant thermodynamical potential which
has to be minimized is the free energy $F$, instead of the energy $E$.
The free energy, the energy, and the entropy $S$ are related through
\begin{equation}
F = E - TS .
\label{eqB1} \end{equation}
In the WK approach $E$ is given by Eq.(\ref{eq34}) and the particle
and kinetic energy densities at finite temperature, for (external
ETF2) nuclear $V_q$ and spin-orbit $W_q$ potentials, for
each kind of particle, read as \cite{BGH,VG}
\begin{eqnarray}
\rho_{{\rm WK},q}^T &=& \frac{1}{2 \pi^2} \bigg( \frac{2mT}{\hbar^2}
\bigg)^{3/2}
\bigg\{ J_{1/2} (\eta_q) + \frac{\hbar^2}{48 m} \bigg[ \frac{\Delta V_q}{T^2}
J_{-3/2}(\eta_q) + \frac{3}{4} \frac{(\nabla V_q)^2}{T^3} J_{-5/2}(\eta_q) 
\bigg]  \nonumber \\ 
&+& \frac{m W_q^2}{2 \hbar^2 T}J_{-1/2}(\eta_q) \bigg\}
\label{eqB2} \end{eqnarray} 
and
\begin{eqnarray} 
\tau_{{\rm WK},q}^T
&=& \frac{1}{2 \pi^2} \bigg( \frac{2mT}{\hbar^2} \bigg)^{5/2}
\bigg\{ J_{3/2} (\eta_q) - \frac{\hbar^2}{48 m} \bigg[5 \frac{\Delta V_q}{T^3}
J_{-1/2}(\eta_q) + \frac{9}{4} \frac{(\nabla V_q)^2}{T^3} J_{-3/2}(\eta_q)
\bigg] \nonumber \\
&+& \frac{5 m W_q^2}{2 \hbar^2 T}J_{1/2}(\eta_q) \bigg\},
\label{eqB3} \end{eqnarray}
where $J_{\nu}(\eta_q)$ are the so-called Fermi
integrals \begin{equation}
J_{\nu}(\eta_q) = \int^{\infty}_{0}  dx \, \frac{x^{\nu}}{1 +
\exp(x-\eta_q)} \label{eqB4} \end{equation}
and $\eta_q=(\mu_q - V_q)/T$ is the fugacity parameter.
In particular, the free energy $F$ for a free Fermi
gas moving in single-particle and spin-orbit potentials is given by
\begin{eqnarray}
F_{{\rm free},q} &=& \mu_q A_q - \frac{1}{2 \pi^2} \bigg(
\frac{2mT}{\hbar^2} \bigg)^{3/2} \int
\bigg\{ d \vec{r} \frac{2}{3} T J_{3/2} (\eta_q) \nonumber \\ 
&-& \frac{\hbar^2}{24 m} \bigg[ 
\frac{\Delta V_q}{T}
J_{-1/2}(\eta_q) + \frac{1}{4} \frac{(\nabla V_q)^2}{T^2} J_{-3/2}(\eta_q)
\bigg] + \frac{m W_q^2}{\hbar^2} J_{1/2}(\eta_q) \bigg\} ,
\label{eqB5} \end{eqnarray}
where $\mu_q$ and $A_q$ are the chemical potential and the particle
number of each kind of nucleon. The final expression of the total free
energy contains in addition to (\ref{eqB5}) the interacting potential
part of the Skyrme-WK energy (\ref{eq34}). The contributions of the
powers of the particle density and its gradients in Eq.~(\ref{eq34})
are expanded in a Taylor series starting from the expression
(\ref{eqB2}) and only the linear terms in $\hbar^2$ are retained.

The ETF2 potentials used in Eqs.(\ref{eqB2}), (\ref{eqB3}) and
(\ref{eqB5}) are obtained at zero temperature, i.e., it is assumed
that they are temperature independent. It should be pointed out that,
even in a fixed external potential, the system starts to evaporate
nucleons as soon as the finite temperature appears \cite{VG} and one
should resort to e.g. a subtraction procedure \cite{S87} to keep the
integrated magnitudes finite and independent of the size of the box in
which they are calculated. However, as far as we are interested in the
$T \to 0$ limit of the free energy, we can safely neglect the effects
from evaporated nucleons because they are negligible below $T \simeq
2$ MeV \cite{BQ}. In such conditions one can consider the
low-temperature expansion \cite{BT} of the free energy and parametrize
it below $T=2$ MeV as $F(T)=E(T=0) - a(T=0) \, T^2$. However, it is to
be noted that the integrals of some terms of the interacting WK free
energy, namely
the ones coming from $(\nabla V_q)^2$ and from the exchange Coulomb
potential, which show a logarithmic divergence at zero temperature,
start to depart from the linear behaviour with $T^2$ below $T=1$ MeV
and bend upwards of the linear curve, similarly to what happens in
Fig.13 where the particular term $\int (\nabla \rho)^2 d \vec{r}$
is plotted as a function of $T^2$.
Thus, we have estimated the WK energy by extrapolating the linear
region in $T^2$ between $T=2$ and $T=1$ MeV\@. As an example, we
display in Fig.14 the results of this procedure for the nuclei
$^{40}$Ca and $^{90}$Zr calculated with the Skyrme T6 force for which
we find $E \simeq -329.3$ and $-766.8$ MeV, respectively.

\newpage

%

\newpage

\begin{table}[h]
\caption{\label{tab1}
 Chemical potential $\mu$ calculated quantally and semiclassically
with  the TF and WK approaches [by inversion of Eq.(\ref{eq20})] as a
function of the accumulated number of fermions occupying the $n=4$
shell in a spherical HO potential. Within this shell the quantal
energy is computed adding a quantum $11 \hbar \omega / 2$ per
fermion to the background energy obtained
filling the previous HO shells. The semiclassical energies are
obtained from Eq.(\ref{eq21}). Both $\mu$ and $E$ are in $\hbar
\omega$ units. Spin degeneracy of each level is assumed.
}

\vspace{1cm}
\begin{tabular}{ccccccc}
\hline
 N & $\mu$(QM) & $\mu$(TF) & $\mu$(WK)
& E(QM) & E(TF) & E(WK) \\
\hline
42 & 5.5 & 5.013 & 5.063 & 161. & 157.919 & 161.092 \\
44 & 5.5 & 5.092 & 5.141 & 172. & 168.024 & 171.296 \\
46 & 5.5 & 5.168 & 5.216 & 183. & 178.284 & 181.653 \\
48 & 5.5 & 5.241 & 5.289 & 194. & 188.693 & 192.159 \\
50 & 5.5 & 5.313 & 5.360 & 205. & 199.248 & 202.809 \\
52 & 5.5 & 5.383 & 5.430 & 216. & 209.945 & 213.599 \\
54 & 5.5 & 5.451 & 5.497 & 227. & 220.780 & 224.526 \\
56 & 5.5 & 5.518 & 5.563 & 238. & 231.750 & 235.587 \\
58 & 5.5 & 5.583 & 5.628 & 249. & 242.851 & 246.778 \\
60 & 5.5 & 5.646 & 5.690 & 260. & 254.080 & 258.096 \\
62 & 5.5 & 5.708 & 5.752 & 271. & 265.434 & 269.539 \\
64 & 5.5 & 5.769 & 5.812 & 282. & 276.912 & 281.103 \\
66 & 5.5 & 5.828 & 5.871 & 293. & 288.510 & 292.787 \\
68 & 5.5 & 5.887 & 5.929 & 304. & 300.225 & 304.588 \\
70 & 5.5 & 5.943 & 5.986 & 315. & 312.056 & 316.504 \\
\hline
\end{tabular}
\end{table}

\begin{table}[h]
\caption{\label{tab2}
Total energies per nucleon of some magic nuclei obtained with the
SkM$^*$ and T6 Skyrme interactions and with the NL3 relativistic mean
field parameter set in several approaches. The Coulomb and spin-orbit
forces are included in the calculations. The column VWK-T refers to
the  empirical ``temperature extrapolation'' method described in
Appendix~2.}

\vspace{1cm}
\begin{tabular}{ccccccccccc}
\hline
& &SkM$^*$&SkM$^*$&T6&T6&T6&NL3&NL3& &\\
\hline
& A & HF & VWK & HF & VWK & VWK-T & H & VWK &\\
\hline
&  16 & $-$7.081 & $-$7.309 & $-$7.026 & $-$7.348 & $-$6.934 & $-$7.282 & $-$6.965 &\\
&  40 & $-$8.126 & $-$8.447 & $-$8.141 & $-$8.428 & $-$8.243 & $-$8.315 & $-$8.265 &\\
&  48 & $-$8.387 & $-$8.565 & $-$8.332 & $-$8.596 & $-$8.336 & $-$8.461 & $-$8.463 &\\
&  90 & $-$8.502 & $-$8.659 & $-$8.514 & $-$8.719 & $-$8.527 & $-$8.603 & $-$8.641 &\\
& 208 & $-$7.779 & $-$7.777 & $-$7.786 & $-$7.828 & $-$7.701 & $-$7.845 & $-$7.817 &\\
\hline
\end{tabular}
\end{table}

\newpage

\begin{figure}
\includegraphics[width=10.5cm]{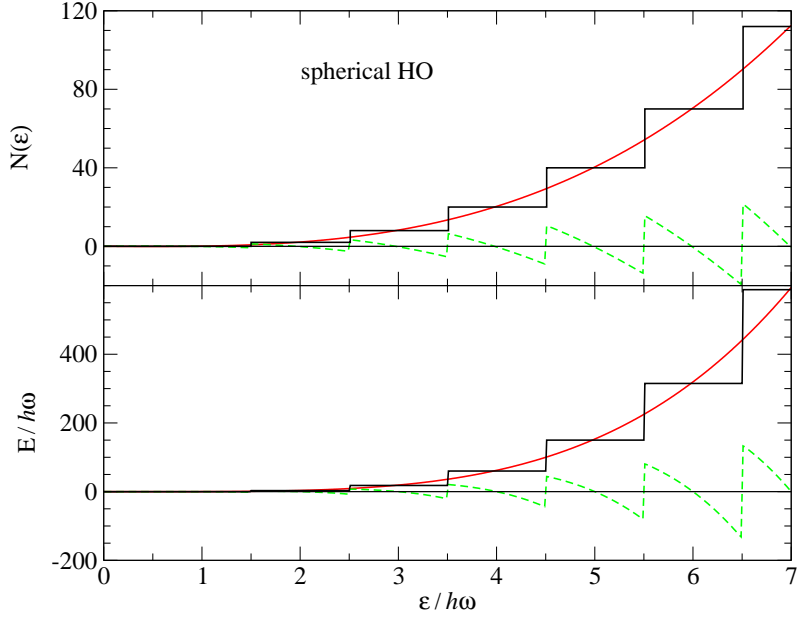}
\caption{(Color online) \label{figure1} Accumulated level density (upper panel) and
total energy (lower panel) with degeneracy 2 for a {\em fixed} spherical
harmonic oscillator potential as a function of the Fermi energy
$\varepsilon$. Staircase, solid, and dashed lines correspond to the
quantal, semiclassical (WK with $\hbar^4$ corrections), and shell
correction (quantal minus semiclassical) values, respectively.
}
\end{figure}

\begin{figure}
\includegraphics[width=9.5cm,angle=270]{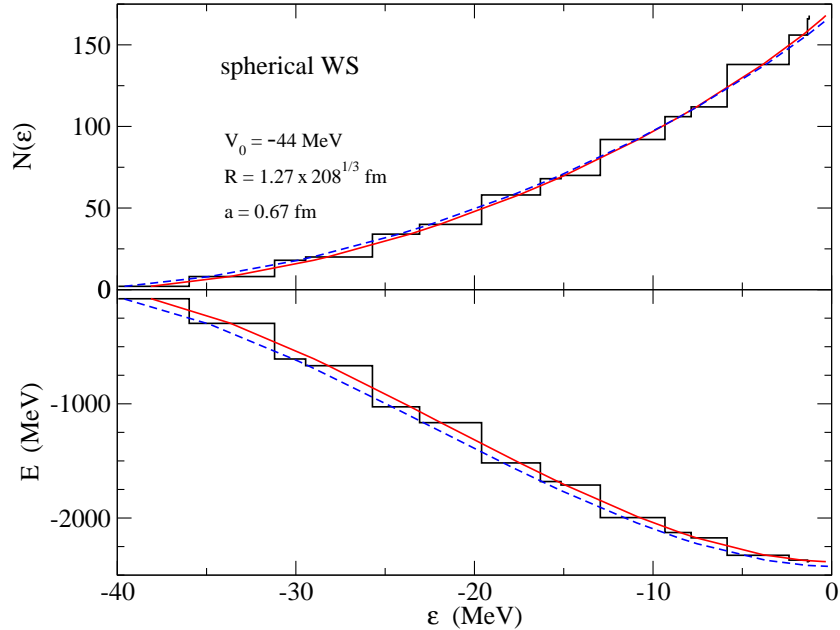}
\caption{(Color online) \label{figure2} Accumulated level density (upper panel) and
total energy (lower panel) with degeneracy 2 for a {\em fixed}
spherical Woods-Saxon potential as a function of the Fermi energy
$\varepsilon$. Staircase lines correspond to the quantal values, while
solid and dashed lines correspond to the semiclassical WK with
$\hbar^2$ corrections and TF results, respectively.}
\end{figure}

\newpage

\begin{figure}
\includegraphics[width=11.5cm,clip=true]{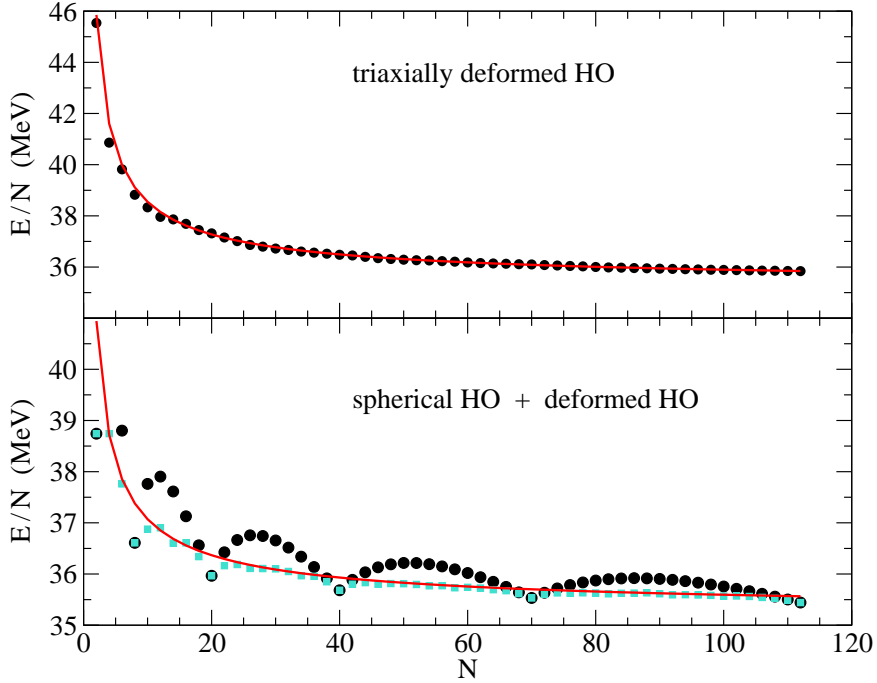}
\caption{(Color online) \label{figure3} Upper panel: 
quantal (dots) and WK (solid line) energy per particle in a strongly
triaxially deformed size-dependent harmonic oscillator potential as a
function of the number of particles. Lower panel: the same as in the
upper panel but for a spherical size-dependent harmonic oscillator
potential (notice that the semiclassical WK curves are different in
the deformed and spherical cases). The squares depict the quantal
energies per particle in the case that the deformation of the harmonic
oscillator potential is optimized leading to maximal binding. Notice
the close agreement with the semiclassical curve obtained for the
spherical harmonic oscillator.} \end{figure}

\begin{figure}
\includegraphics[width=10.5cm,angle=0,clip=true]{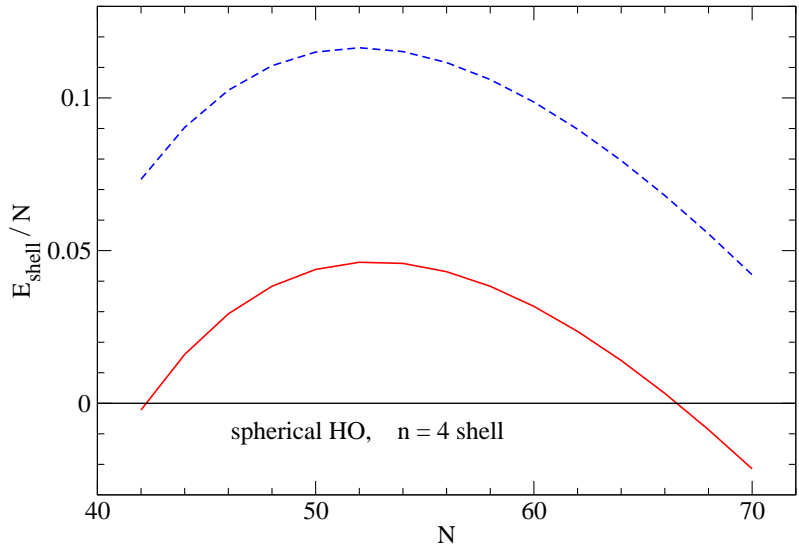}
\caption{(Color online) \label{figure4} WK2 (solid line) and TF (dashed line) shell
energies, per particle, defined as the difference between the quantal
and semiclassical energies (in $\hbar \omega$ units) filling the $n=4$
shell of a fixed spherical harmonic oscillator potential as a function
of the number of the particles in the shell. The $\hbar^4$-corrections are 
indistinguishable from the solid VWK2-line.}
\end{figure}

\newpage

\begin{figure}
\includegraphics[width=12.0cm]{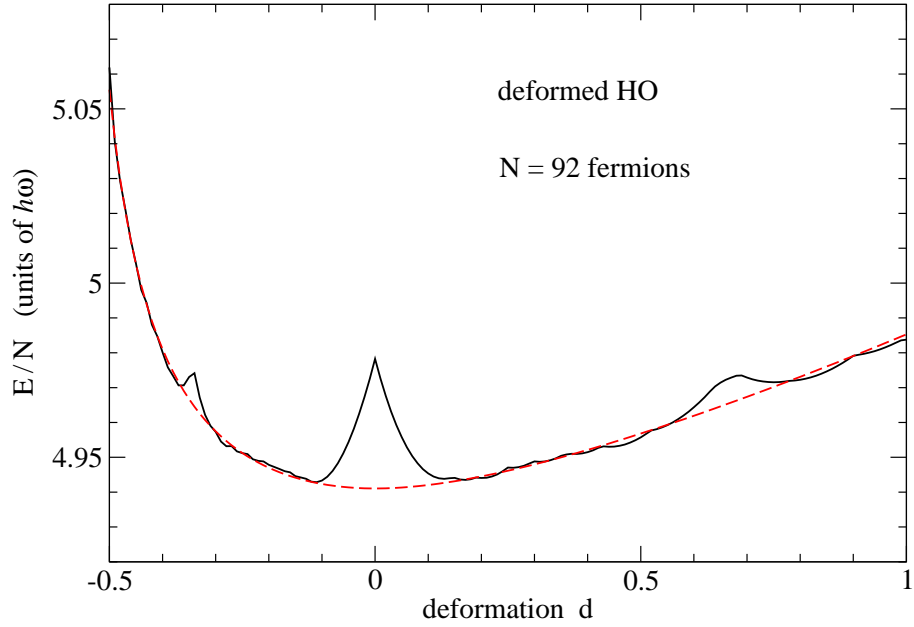}
\caption{(Color online) \label{figure5} Quantal (solid line) and WK (dashed line) 
values of the energy per particle of a set of 92 fermions submitted to
a triaxilly deformed HO potential as a function of the deformation $d$
[Eqs.(\ref{eq25}) and (\ref{eq25a})]. Spin degeneracy is included.}
\end{figure}

\newpage 

\begin{figure}
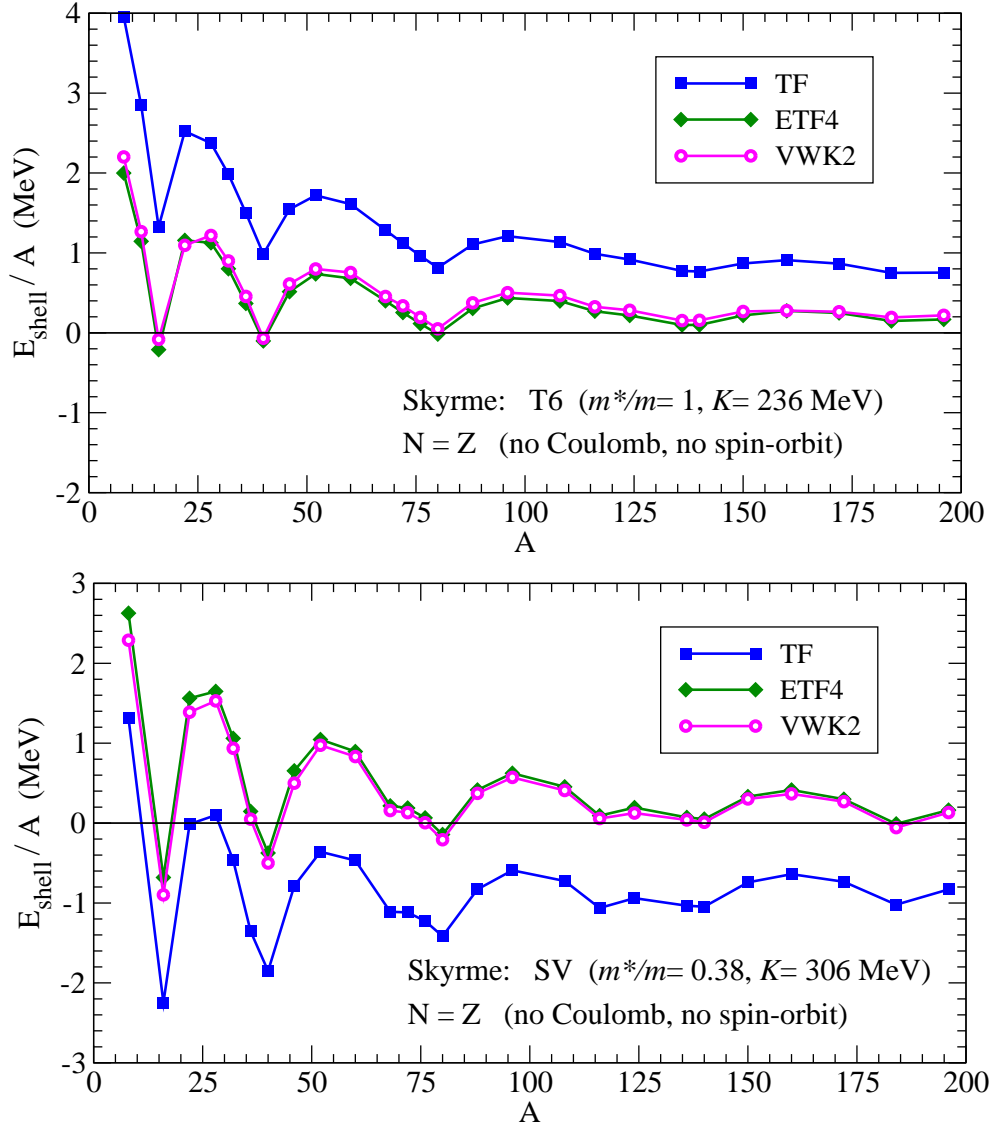

\includegraphics[width=13cm,angle=0]{vwkf06_shell_T6_symm.eps}
\vspace*{0.2cm}

\includegraphics[width=13cm,angle=0,clip=true]{vwkf06_shell_SV_symm.eps}
\caption{(Color online) \label{figure6} 
Upper panel: Shell correction $E_{\rm shell} = E_{\rm HF} - E_{\rm
semicl}$ in the TF, ETF4, and VWK2 approaches for symmetric uncharged
nuclei without spin-orbit force as a function of the mass number $A$
calculated with the Skyrme force T6 \cite{T6}.
Lower panel: the same for the Skyrme force SV \cite{Sk5}.} 
\end{figure}

\newpage

\begin{figure}
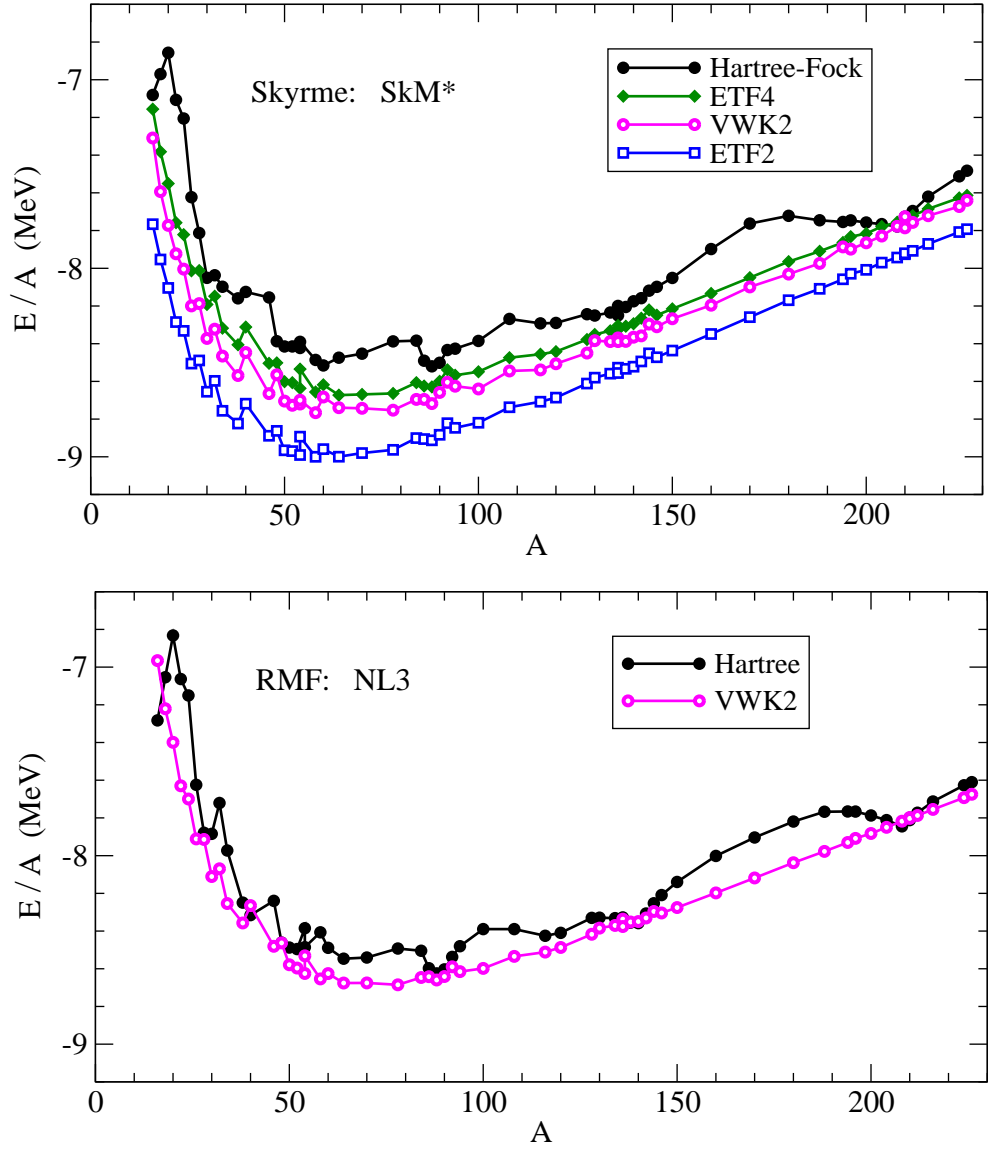

\includegraphics[width=13.cm]{vwkf07_eoa_SKM.eps}
\vspace*{0.4cm}

\includegraphics[width=13.cm,clip=true]{vwkf07_eoa_NL3.eps}
\caption{(Color online) \label{figure11} Energy per nucleon of
$\beta$-stable spherical nuclei along the periodic table calculated with the
non-relativistic Skyrme force SkM$^*$ (upper panel) and with
relativistic mean field with parameter set NL3 (lower
panel).} \end{figure}
 
\newpage

\begin{figure}
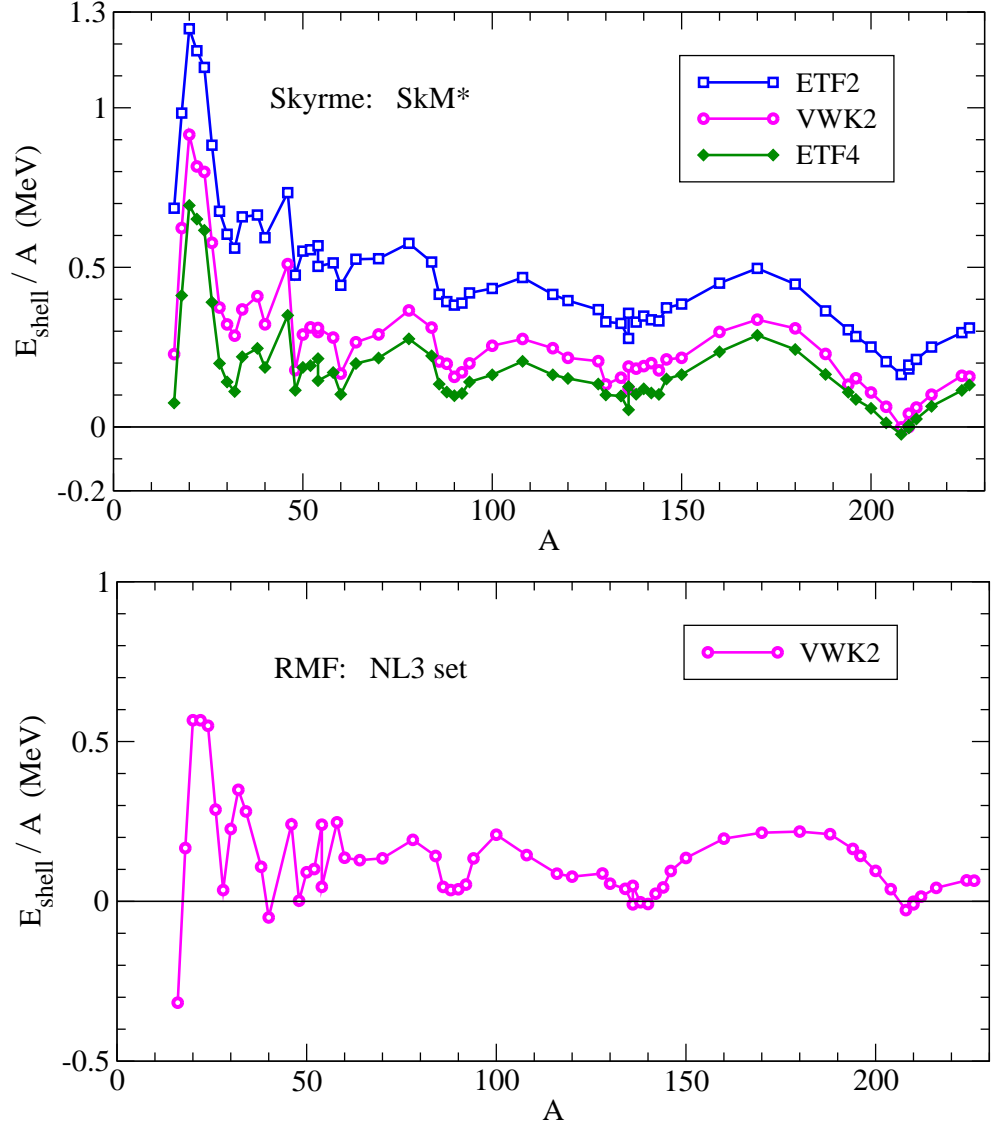

\includegraphics[width=13cm]{vwkf08_shell_SKM.eps}
\vspace*{0.2cm}

\includegraphics[width=13cm,clip=true]{vwkf08_shell_NL3.eps}
\caption{(Color online) \label{figure13}
Shell correction at VWK2 level of $\beta$-stable spherical nuclei
as a function of the mass number $A$ calculated with
the Skyrme force SkM$^*$ (upper panel) and using 
relativistic mean field with parameter set NL3 (lower panel).}
\end{figure}

\newpage

\begin{figure}
\includegraphics[width=12.cm]{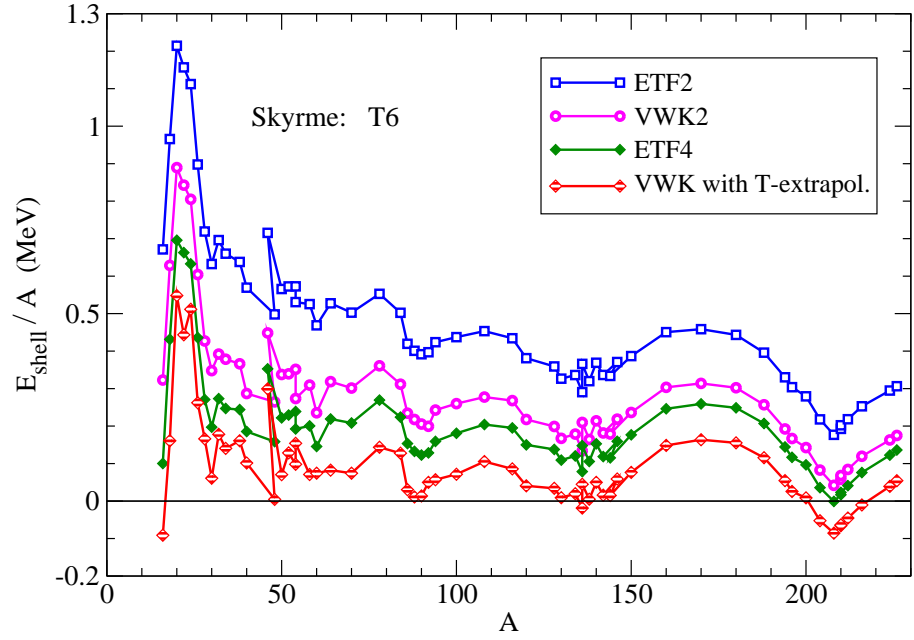}
\caption{(Color online) \label{figure15}
Shell energies per nucleon along the $\beta$-stability line computed with the 
Skyrme force T6 in several approaches, including the empirical
temperature extrapolation method described in Appendix~2.}
\end{figure}

\newpage

\begin{figure}
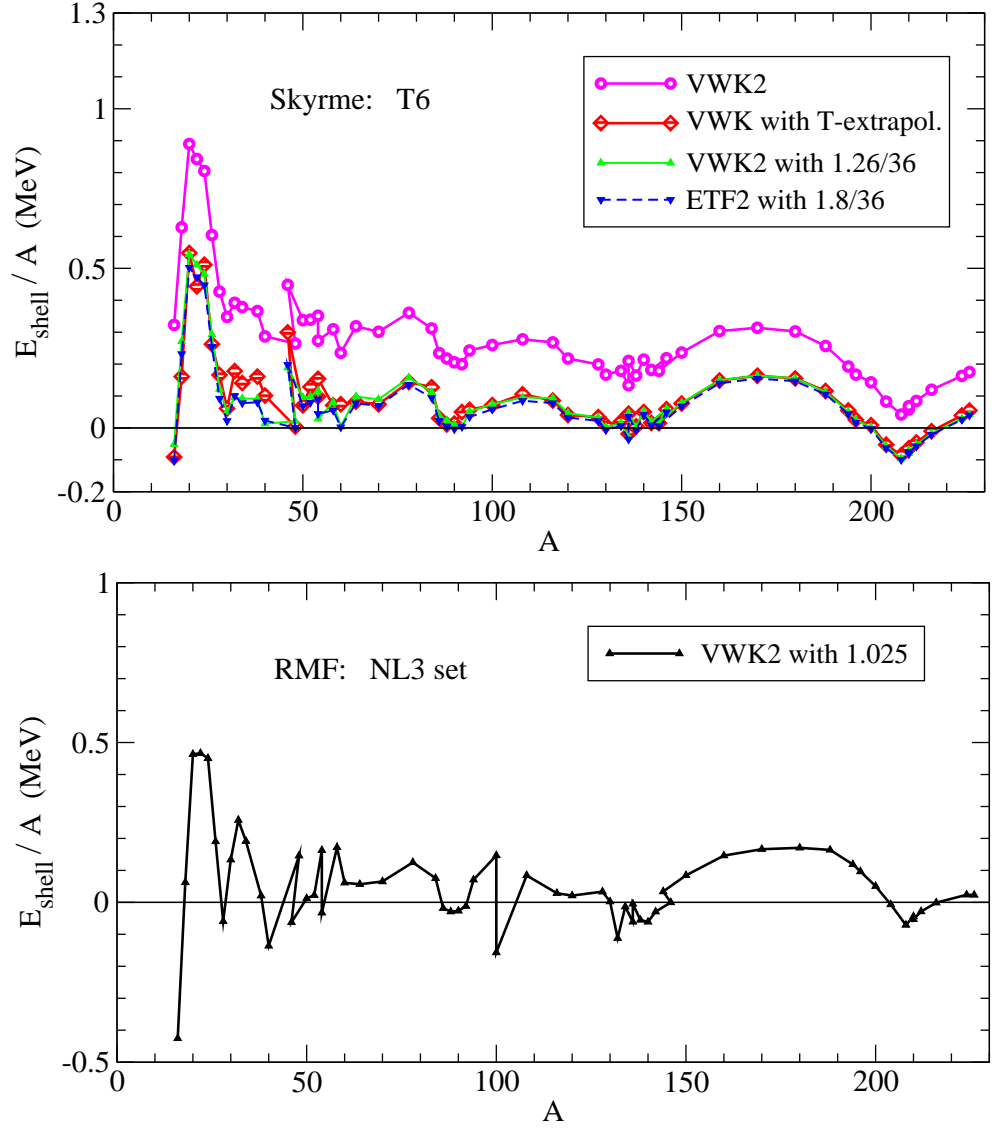

\includegraphics[width=13cm]{vwkf10_shell_T6_with_T-extrapol_and_Fudge.eps}
\vspace*{0.2cm}

\includegraphics[width=13cm,clip=true]{vwkf10_shell_NL3_with_Fudge.eps}
\caption{(Color online) \label{figure16}
Upper panel: shell energies per nucleon along the $\beta$-stability
line for spherical nuclei computed with the
Skyrme force T6 using the VWK2 approach and the different empirical methods
described in the text. Lower panel: same for the RMF parameter
set NL3. In this case the doubly magic, non $\beta$-stable nuclei
$^{100}$Sn and $^{132}$Sn have been added for more complete
information.} \end{figure}

\newpage

\begin{figure}
\includegraphics[width=11.cm]{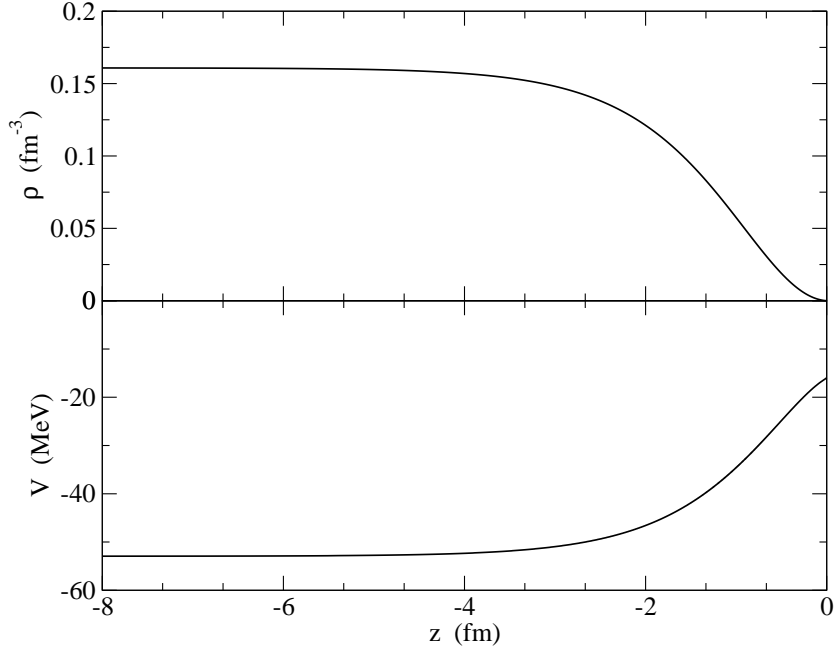}
\caption{\label{figure20}
Thomas-Fermi density (upper panel) and single-particle potential
(lower panel) profiles
in half infinite symmetric nuclear matter obtained using the T6 \cite{T6} 
force. The fact that the potential enters the classical turning point
with a horizontal tangent cannot be seen on the scale of the graph.}
 \end{figure}

\begin{figure}
\includegraphics[width=11.cm]{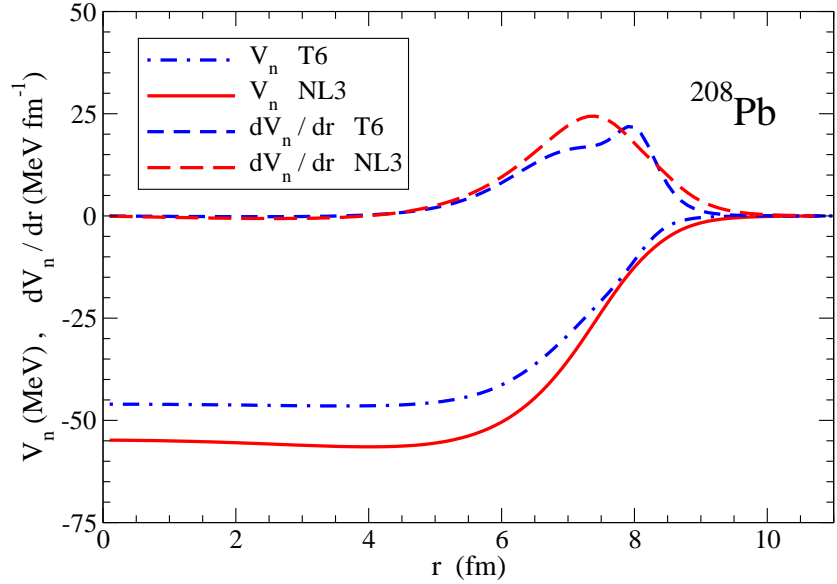}
\caption{(Color online) \label{figure21}
Neutron single-particle potential and its derivative for the nucleus 
$^{208}$Pb obtained with the T6 Skyrme interaction in ETF2 and with
the relativistic NL3 parameter set in TF approximation.}
\end{figure}

\newpage

\begin{figure}
\includegraphics[width=11.cm]{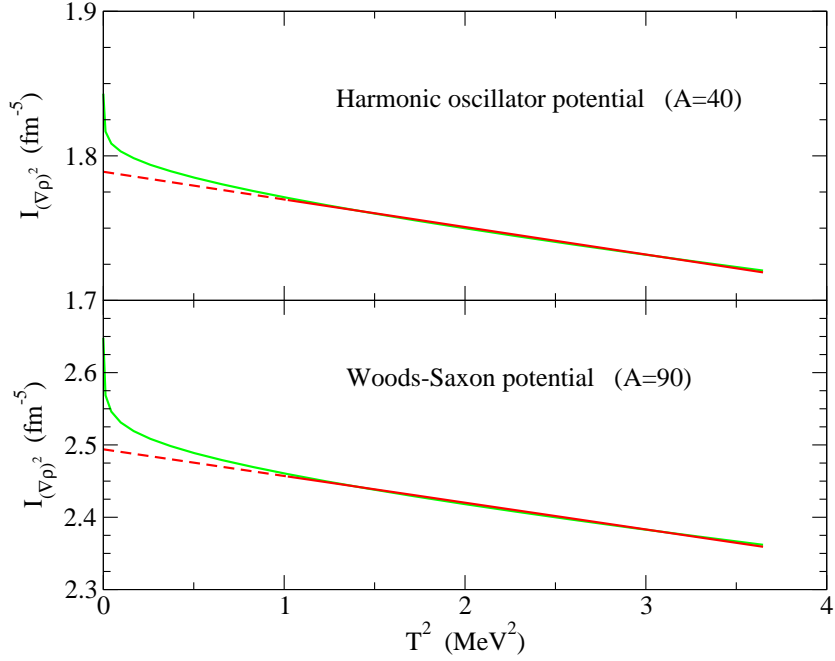}
\caption{(Color online) \label{figure23}
Extrapolation to $T=0$ of the WK value of $\int (\nabla \rho)^2 d 
\vec{r}$ obtained from HO (upper panel) and  WS (lower panel) 
potentials.} 
\end{figure}

\begin{figure}
\includegraphics[width=11.cm]{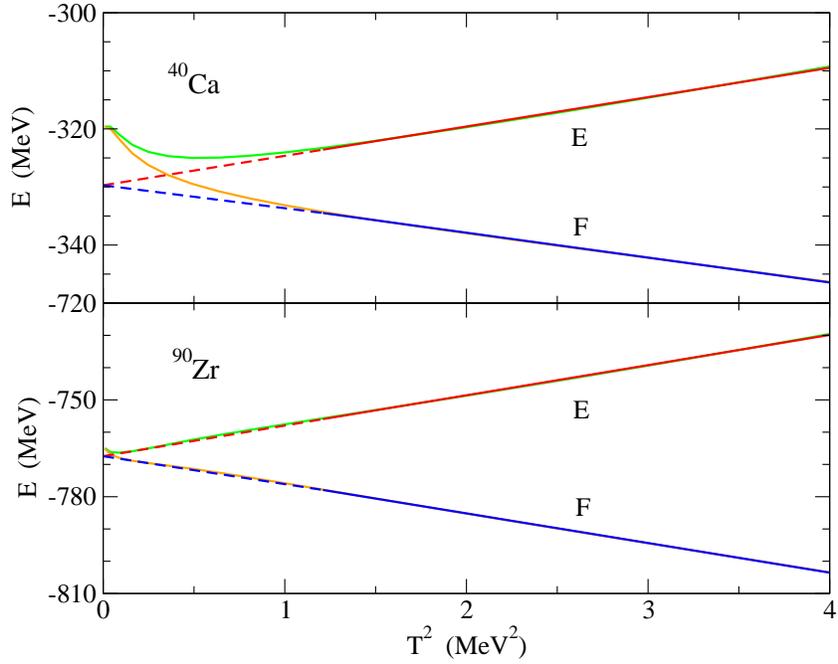}
\caption{(Color online) \label{figure24}
Extrapolation to $T=0$ of the Skyrme energy ($E$) and free energy
($F$) corresponding to the T6 force obtained with the thermal WK
method on top of the self consistent ETF2 potential
for $^{40}$Ca (upper panel) and $^{90}$Zr (lower panel).}
\end{figure}

\end{document}